\documentstyle[aps,prb,eqsecnum]{revtex}

\newcommand{\r}{{\it r}}
\newcommand{\s}{{\it s}}
\newcommand{\k}{{\it k}}
\newcommand{\iit}{{\it i}}
\newcommand{\beq}{\begin{equation}}
\newcommand{\eeq}[1]{\label{#1} \end{equation}}
\newcommand{\diag}{{\rm diag\,}}
\newcommand{\tr}{{\rm tr\,}}

\newcommand{\trg}{{\rm trg\,}}
\newcommand{\detg}{{\rm detg\,}}

\newcommand{\beqar}{\begin{eqnarray}}
\newcommand{\eeqar}[1]{\label{#1} \end{eqnarray}}

\begin{document}

\title{Recursive Construction for a Class of Radial 
       Functions II --- Superspace}
\author{Thomas Guhr\thanks{e-mail: guhr@daniel.mpi-hd.mpg.de}
        and
        Heiner Kohler\thanks{e-mail: kohler@daniel.mpi-hd.mpg.de}\\
        Max Planck Institut f\"ur Kernphysik,
        Postfach 103980,
        69029 Heidelberg,
        Germany}






\maketitle

\begin{abstract}
  We extend the recursion formula for matrix Bessel functions, which
  we obtained previously, to superspace.  It is sufficient to do this
  for the unitary orthosymplectic supergroup.  By direct computations,
  we show that fairly explicit results can be obtained, at least up to
  dimension $8\times 8$ for the supermatrices. Since we introduce a
  new technique, we discuss various of its aspects in some detail.
\end{abstract}

\section{Introduction}
\label{SS}

In a previous work, we studied properties of matrix Bessel functions
in ordinary space~\cite{GK1}.  Here, we generalize these investigations
to superspace. For the introductory remarks and the mathematical and
physical background relevant for the ordinary space, and also relevant
as the basis for the present study, we refer the reader to
Ref.~\cite{GK1}.

In mathematics, supersymmetry was pioneered by Berezin~\cite{BER} and,
in particular group theoretical aspects, by Kac~\cite{KAC1,KAC2}.  The
theory of non--linear $\sigma$ models in spaces of supermatrix fields
was developed in physics of disordered systems by
Efetov~\cite{Efetov,EfeB}.  Verbaarschot, Weidenm\"uller and
Zirnbauer~\cite{VZ,VWZ} used his approach to study models in Random
Matrix Theory. In Ref.~\cite{TG}, the first supersymmetric
generalization of the Itzykson--Zuber integral~\cite{IZ} was given. In
Ref.~\cite{GGT}, Gelfand--Tzetlin coordinates~\cite{GT} were
constructed for the unitary supergroup. Extending
Shatashvili's~\cite{SLS} method, the supersymmetric Itzykson--Zuber
integral was also rederived in Ref.~\cite{GGT} in its most general
form. Using the techniques of Ref.~\cite{TG}, such a calculation was
also performed in Ref.~\cite{AMU}.

{}From a mathematical viewpoint, Efetov's work~\cite{Efetov} is the
basis for a harmonic analysis in certain supersymmetric coset spaces,
the Efetov spaces, which are relevant for the non--linear $\sigma$
models.  In the full superspaces, a technique involving convolution
integrals and ingredients of the corresponding harmonic analysis was
introduced in Ref.~\cite{GW}.  In the Efetov spaces, the theory of
harmonic analysis, in both its mathematical and physical aspects, was
developed by Zirnbauer~\cite{Zirn1} and was applied to disordered
systems in Refs.~\cite{Zirn2,Mirlin}.  In the present contribution, we
do not focus on the Efetov spaces, rather we address the full
supergroup spaces.  The supersymmetric Itzykson--Zuber
integral~\cite{TG} and its application in Ref.~\cite{GW} is the
simplest example of a supermatrix Bessel function appearing in this
kind of harmonic analysis.

The matrix Bessel functions in superspace find direct application in
Random Matrix Theory. For general reviews, see
Refs.~\cite{Haake,MEH,GMGW}. In Ref.~\cite{GUH4} it was shown that
they are the kernels for the supersymmetric analogue of Dyson's
Brownian Motion.

The paper is organized as follows: In Sec.~\ref{supBes}, we introduce
the supermatrix Bessel functions and collect basic definitions and
notations. In Sec.~\ref{subs321}, we extend the recursion formula of
Ref.~\cite{GK1} to superspaces. Since it is one of our goals to
demonstrate that explicit results for supermatrix Bessel functions can
indeed be obtained, we present, in some detail, such calculations for
certain supermatrix Bessel functions in Secs.~\ref{subs322}
and~\ref{subs323}, respectively. The asymptotics and the normalization
are discussed in Sec.~\ref{asym}. We briefly comment on applications
in Sec.~\ref{appl} and we summarize and conclude in Sec.~\ref{sum4}.
Various calculations are shifted to the appendix.

\section{Supermatrix Bessel Functions}
\label{supBes}

Similar to ordinary spaces~\cite{GK1}, the superunitary case,
i.e.~integration over the supergroup ${\it U}(k_1/k_2)$, is the
simplest one.  As it was already discussed in detail in
Refs.~\cite{TG,GGT,GUH4}, we refrain from reconsidering it here.
Thus, it turns out that we may restrict ourselves to the supermatrix
Bessel function of the unitary orthosymplectic group ${\it
  UOSp}(k_1/2k_2)$. As discussed by Kac~\cite{KAC1,KAC2}, the
supergroups ${\it U}(k_1/k_2)$ and ${\it UOSp}(k_1/2k_2)$ exhaust
almost all classical compact supergroups, apart from some exotic exceptions
which are of little relevance for applications. Thus, the integral we
have to deal with is given by
\begin{equation}
\Phi_{k_12k_2}(s,r) \ = \ \int_{u\in {\it UOSp}(k_1/2k_2)}  
                      \exp\left(i \trg u^{-1}sur\right) d\mu(u) \ ,
\label{321}
\end{equation}
where $d\mu(u)$ is the invariant measure. The arguments of the
function~(\ref{321}) are the diagonal matrices $s=\diag(\sqrt{c}
s_1,\sqrt{-c} s_2)$ and $r=\diag(\sqrt{c} r_1,\sqrt{-c} r_2)$.  Here,
we use Wegner's notation~\cite{WEG1} and introduce the label $c=\pm 1$
to distinguish the two possible forms.  We will return to this issue.
The matrices $s_1$, $s_2$ and $r_1$, $r_2$ are given by
\begin{eqnarray}
s_1=\diag(s_{11},s_{21},\ldots,s_{k_11})& , \quad &
s_2=\diag(s_{12}1_2,\ldots,s_{k_22}1_2) \ , \nonumber\\
r_1=\diag(r_{11},r_{21},\ldots,r_{k_11})& , \quad &
r_2=\diag(r_{12}1_2,\ldots,r_{k_22}1_2) \ .
\label{323}
\end{eqnarray}
There is a twofold degeneracy in $s_2$ and $r_2$, because the
matrix $u^{-1}su$ or, equivalently, $uru^{-1}$ has to be
a {\it real Hermitean} supermatrix~\cite{WEG1} of the form 
\begin{equation}
\sigma=\left[\matrix{\sqrt{c} \sigma^{(R)}&\sigma^{(A)\dagger}\cr
                \sigma^{(A)}&\sqrt{-c}\sigma^{(HSd)}}\right]
       \quad,\qquad c=\pm 1 \ .
\label{diffs6}
\end{equation}
The matrices $\sigma^{(R)}$ and $\sigma^{(HSd)}$ have ordinary
commuting entries, i.e.~bosons, they are real symmetric and Hermitean
self--dual, respectively. The matrix $\sigma^{(A)}$ has anticommuting
or Grassmann entries, i.e.~fermions, and is of the form
\begin{equation}
\sigma^{(A)}=[\sigma_1^{(A)},\ldots,\sigma_{k_1}^{(A)}]\ ,\qquad
\sigma^{(A)}_i=\left[\matrix{\sigma^{(A)}_{1i}\cr
                      \sigma^{(A)*}_{1i}\cr
                      \vdots\cr
                      \sigma^{(A)}_{k_2i}\cr
                      \sigma^{(A)*}_{k_2i}}\right]\quad .
\label{diffs7}
\end{equation}     
We can now appreciate the meaning of the parameter $c$ which enters
the definition (\ref{diffs6}) of the real Hermitean matrices.  For
$c=1$, it yields the real symmetric and for $c= -1$ the Hermitean
self--dual matrix as boson--boson block, and vice versa for the
fermion--fermion block. In the framework of Random Matrix Theory,
we find the supermatrices corresponding to the Gaussian Orthogonal
Ensemble (GOE) for $c=+1$ and those for the Gaussian Symplectic
Ensemble (GSE) for $c=-1$.

The infinitesimal volume element is given by
\begin{equation}
d[\sigma]= \prod_{i=1}^{k_1}\prod_{j=1}^{k_2}d\sigma^{(A)*}_{ij}d\sigma^{(A)}_{ij}
           \prod_{i<j}d\sigma^{(R)}_{ij}\prod_{i=1}^{k_1}d\sigma^{(R)}_{ii}
           \prod_{i<j}d[\sigma^{(HSd)}_{ij}]
           \prod_{i=1}^{k_2}d\sigma^{(HSd)}_{ii}
           \ ,
\label{diffs7a}
\end{equation} 
where $d[\sigma^{(HSd)}_{ij}]$ is the product of the differentials of
all independent elements of the quaternion $\sigma^{(HSd)}_{ij}$.  

The supermatrix Bessel functions~(\ref{321}) are eigenfunctions of a
wave equation in the curved space of the eigenvalues $s$ or $r$.  As
in the ordinary case, a supermatrix gradient $\partial/\partial\sigma$
is introduced and the Laplace operator is defined by
\begin{equation}
\Delta \ = \ \trg\left(\frac{\partial}{\partial\sigma}\right)^2 \ . 
\label{diffs11}
\end{equation}
The plane waves $\exp(i\trg\sigma\rho)$ are the eigenfunctions,
i.e.~we have
\begin{equation}
\Delta \exp(i\trg\sigma\rho) \ = \ -\trg\rho^2
\exp(i\trg\sigma\rho) \ .
\label{323a}
\end{equation}
Here, both $\sigma$ and $\rho$ are real Hermitean.  As in ordinary
space, the supermatrix Bessel functions are obtained by averaging over
the angular coordinates, i.e.~over the diagonalizing group . The
Laplacean commutes with the average and we arrive at the
differential equation
\begin{equation}
\Delta_s \Phi_{k_12k_2}(s,r) = -\trg r^2 \Phi_{k_12k_2}(s,r) \ ,
\label{323b}
\end{equation}
where the radial part of of the Laplacean~(\ref{diffs11}) reads
\begin{equation}
\Delta_s=\frac{1}{\widetilde{B}_{k_1k_2}^{(c)}(s)}\left(
           \sum_{p=1}^{k_1}
           \frac{\partial}{\partial s_{p1}}{\widetilde{B}_{k_1k_2}^{(c)}(s)}
           \frac{\partial}{\partial s_{p1}}
           +\frac{1}{2}\sum_{p=1}^{k_2}
           \frac{\partial}{\partial s_{p2}}{\widetilde{B}_{k_1k_2}^{(c)}(s)}
           \frac{\partial}{\partial s_{p2}}\right) \ .
\label{234}
\end{equation}
The Jacobian or Berezinian is given by~\cite{GUH4}
\begin{equation}
\widetilde{B}_{k_1k_2}^{(1)}(s)=\frac{|\Delta_{k_1}(s_1)
                 |\Delta_{k_2}^4(is_2)}
            {\prod_{p=1}^{k_1}\prod_{q=1}^{k_2}(s_{p1}-is_{q2})^2}\quad ,
        \quad
\widetilde{B}_{k_1k_2}^{(-1)}(s)=\frac{|\Delta_{k_1}(is_1)
                 |\Delta_{k_2}^4(s_2)}
                 {\prod_{p=1}^{k_1}\prod_{q=1}^{k_2}(is_{p1}-s_{q2})^2}\ .
\label{325}
\end{equation}
One easily convinces oneself that $\Delta_s$ depends on $c$ only
through a factor $\sqrt{c}$.  Thus, without loss of generality, we set
$c=1$ and omit the index $c$.

At this point, an important comment is in order. The normalization in
ordinary space according to Eqs.~(3.17) in Ref.~\cite{GK1},
$\Phi_N^{(\beta)}(x,0)=1$ and $\Phi_N^{(\beta)}(0,k)=1$, do not carry
over to the supersymmetric case. This is due the fact that the volume
of some supergroups is zero \cite{BER} resulting in the vanishing of
$\Phi_{k_12k_2}(0,s)$ for certain values of $k_1$ and $k_2$.  This
collides with the normalization of the plane waves (\ref{323a}) to
unity at the origin. The reason of this contradiction is a well known
phenomenon in superanalysis.  In going from Cartesian to angle
eigenvalue coordinates, one has to add additional terms to the measure
to preserve the symmetries of the original integral. These are called
Efetov--Wegner--Parisi--Sourlas terms in physical literature.  A
full--fledged mathematical theory of these boundary terms was given by
Rothstein \cite{ROT}.

To solve this normalization problem, we use the following strategy.
First, we evaluate the supermatrix Bessel functions without taking
care of the normalization. We just multiply the integrals with a
normalization constant $\widehat{G}_{k_12k_2}$. Having done the
integrals, we determine the normalization by comparing the asymptotics
of the supermatrix Bessel function for large arguments with the
Gaussian integral.

\section{Supersymmetric Recursion Formula}
\label{subs321}

We extend the recursion formula in ordinary space~\cite{GK1} to
superspace. After stating the result in Sec.~\ref{res},
we present the derivation and the calculation of the invariant
measure in Secs.~\ref{deriv} and~\ref{inv}, respectively.

\subsection{Statement of the Result}
\label{res}  

Let $\Phi_{k_12k_2}(s,r)$ be defined through the group integral in
Eq.~(\ref{321}). It has two diagonal matrices defined as in
Eq.~(\ref{323}) as arguments. It can be calculated iteratively
by the {\it recursion formula}
\begin{equation}
\Phi_{k_12k_2}(s,r) \ = \widehat{G}_{k_12k_2}\; 
         \int d\mu(s^\prime,s) \, 
         \exp\left(i(\trg s - \trg s^\prime)r_{11} \right) \, 
         \Phi_{(k_1-1)2k_2}(s^\prime,\widetilde{r})\quad ,
\label{susyrec}
\end{equation}
where $\Phi_{(k_1-1)2k_2}^{(\beta)}(s^\prime,\widetilde{r})$ is the
group integral (\ref{321}) over ${\it UOSp}\left((k_1-1)/2k_2)\right)$
and $\widehat{G}_{k_12k_2}$ is a normalization constant, see the
previous section and Sec.~\ref{asym}.  As in the ordinary
case~\cite{GK1}, the coordinates $s^\prime$ are {\it radial}
Gelfand--Tzetlin coordinates. Again, they are different from the {\it
  angular} Gelfand--Tzetlin coordinates which will be discussed
elsewhere~\cite{GK2}.  We also introduced the diagonal matrix
\begin{equation}
\widetilde{r}=\diag(r_{21},\ldots,r_{k_11},ir_2)
              =\diag(\widetilde{r}_1,i\widetilde{r}_2)
\label{susyre2}
\end{equation} 
such that  $r=\diag(r_{11},\widetilde{r})$ and the diagonal matrix
\begin{equation}
s^\prime=\diag(s_{11}^\prime,\ldots,s_{(k_1-1)1}^\prime,is_2^\prime)
        =\diag(s_1^\prime,is_2^\prime)\quad .
\label{susyre2a}
\end{equation}
The invariant measure reads
\begin{eqnarray}
d\mu(s^\prime,s)& = &2^{k_2+1}\mu_B(s_1^\prime,s_1)\mu_F(s_2^\prime,s_2)
                            \mu_{BF}(s^\prime,s)d[\xi^\prime]d[s_1^\prime]\cr
\mu_B(s_1^\prime,s_1)& = & \frac{\Delta_{k_1}(s_1^\prime)}
                                {\sqrt{-\prod_{p=1}^{k_1}
                                         \prod_{q=1}^{k_1-1}
                                         (s_{p1}-s_{q1}^\prime)}}\cr
\mu_F(s_2^\prime,s_2)& = & \frac{\Delta_{k_2}^4(is_2^\prime)}
                               {\prod_{p=1}^{k_2}\prod_{q=1}^{k_2}
                                            (is_{p2}-is_{q2}^\prime)^2}\cr
\mu_{BF}(s^\prime,s)& = & \frac{\prod_{p=1}^{k_1}\prod_{l=1}^{k_2}
                                \prod_{q=1}^{k_1-1}(is_{l2}^\prime-s_{p1})
                                                   (is_{l2}-s_{q1}^\prime)}
                              {\prod_{p=1}^{k_1-1}\prod_{l=1}^{k_2}
                               (is_{l2}^\prime-s_{p1}^\prime)^2}\ .
\label{ss7}
\end{eqnarray}
Here, we have introduced the differentials
\begin{equation}
d[\xi^\prime]=\prod_{p=1}^{k_2}d\xi_p^{\prime*}d\xi_p^\prime \quad ,\quad
d[s_1^\prime]=\prod_{p=1}^{k_1-1}ds_{p1}^\prime\ .
\label{ss7a}
\end{equation}
The domain of integration for the bosonic variables is compact and
given by
\begin{equation}
s_{p1} \ \le \ s_{p1}^\prime \ \le \ s_{(p+1)1} \ , 
        \qquad p=1,\ldots,(k_1-1) \ .
\label{ss7b}
\end{equation}
The fermionic eigenvalues $is_{p2}^\prime$ are related to Grassmann
variables $\xi_p^\prime$ and $\xi_p^{\prime*}$ through 
\begin{equation}
|\xi_p^\prime|^2 \ = \ is_{p2}^\prime-is_{p2}  \ .
\label{ss6}
\end{equation}
The Jacobian or Berezinian consists of three parts. One of them,
$\mu_B(s_1^\prime,s_1)$, depends only on bosonic eigenvalues and one,
$\mu_F(s_1^\prime,s_1)$, only on fermionic eigenvalues, i.~e.~only on
Grassmann variables. The third part mixes commuting and anticommuting
integration variables. To underline once more the difference between
radial and angular Gelfand--Tzetlin coordinates which is also present
in superspace, we mention that the radial measure~(\ref{ss7}) is 
quite different from the angular one~\cite{GK2}.

As in ordinary space, the recursion formula is an exact map of the
group integration onto an iteration exclusively in the radial space.
Having done the iteration on the first $k_1$ levels, we have treated
all Grassmann variables. Thus, in the integrand, we are left with the
matrix Bessel function $\Phi_{k_2}^{(4)}(-i2s_2^{(k_1-1)},r_2)$ for
${\it USp}(2k_2)$ in ordinary space~\cite{GK1},
\begin{eqnarray}
\lefteqn{\Phi_{k_12k_2}(s,r) \ = \
       \int \prod_{n=1}^{k_1-1} d\mu(s^{(n)},s^{(n-1)})} \nonumber\\
    & & \exp\left(i(\trg s^{(n-1)}\!-\!\trg s^{(n)})r_{n1}\right)
        \exp\left(is_{11}^{(k_1-1)}r_{k_11}\right) \, \Phi_{k_2}^{(4)}
       (-i2s_2^{(k_1-1)},r_2) \ .
\label{ss6aa}
\end{eqnarray}
We have set $s=s^{(0)}$ and $s^\prime=s^{(1)}$.  It is worthwhile to
notice that the radial Gelfand--Tzetlin coordinates have a highly
appreciated and valuable property: The Grassmann variables only appear
as moduli squared in the integrand.  Thus, the number of integrals
over anticommuting variables is only {\it half} the number of the
independent Grassmann variables.  Moreover, advantageously, the
exponential is a simple function of the integration variables. Thus,
we may conclude that the radial Gelfand--Tzetlin coordinates are the
natural coordinates of the matrix and the supermatrix Bessel
functions, because their intrinsic features are reflected.

\subsection{Derivation}  
\label{deriv}

All crucial steps needed for the derivation of the supersymmetric
recursion formula~(\ref{susyrec}) carry over from the ordinary
recursion formula in Ref.~\cite{GK1}. We order the columns of the
matrix $u\in {\it UOSp}(k_1/2k_2)$ in the form
$u=[u_{1}\, u_{2}\, \cdots\, u_{k_1}\, u_{k_1+1}\, \cdots\, u_{k_1+k_2}]$.
We also introduce a rectangular matrix
$b=[u_{2}\, \cdots\, u_{k_1}\, u_{k_1+1}\, \cdots\, u_{k_1+k_2}]$ such that 
$u=[u_{1}\, b]$. Analogously to the ordinary case, we have 
\begin{eqnarray}
b^\dagger b &=& 1_{(k_1-1)2k_2} \nonumber\\
b b^\dagger &=& \sum_{p=2}^{k_1} u_pu_p^\dagger \, + \,
                \sum_{p=k_1+1}^{k_1+k_2} u_pu_p^\dagger 
           \ = \ 1_{k_12k_2}-u_{1}u_{1}^\dagger \ .
\label{der1}
\end{eqnarray}
We define the square matrix $\widetilde{\sigma}=b^\dagger s b$ and
rewrite the trace in the exponent as
\begin{equation}
\trg u^\dagger su r  \ = \ \trg\widetilde{\sigma}\widetilde{r}
                                   \, + \, \sigma_{11}
                                   r_{11}\quad ,
\label{der2}
\end{equation}
with $\sigma_{11}=u_{1}^\dagger su_{1}$.  Similarly to the ordinary
case, the first term of the right hand side of Eq.~(\ref{der2})
depends on the last $k_1-1+k_2$ columns $u_p$ collected in $b$ and the
second term depends only on $u_{1}$.  Thus, it is useful to decompose
the invariant measure,
\begin{equation}
d\mu(u) \ = \ d\mu(b) \, d\mu(u_{1}) \ ,
\label{der3}
\end{equation}
and to write Eq.~(\ref{321}) in the form
\begin{equation}
\Phi_{k_12k_2}(s,r) \ = \ \int d\mu(u_{1}) \, 
                          \exp(i\sigma_{11} r_{11}) \,
                          \int d\mu(b) \, 
                          \exp(i\trg\widetilde{\sigma}\widetilde{r}) 
                          \ .
\label{der4}
\end{equation}
Since the coordinates $b$ are locally orthogonal to $u_{1}$, the
measure $d\mu(b)$ also depends on $u_{1}$.

We now generalize the radial Gelfand--Tzetlin coordinates introduced
in \cite{GK1} for the ordinary spaces to the superspace. Naturally,
the projector reads $(1_{k_12k_2}-u_{1}u_{1}^\dagger)$ 
and we have the defining equation
\begin{equation}
(1_{k_12k_2}-u_1u_1^\dagger) \, s \, 
(1_{k_12k_2}-u_1u_1^\dagger) \, e_p^\prime \ = \
                            s_p^\prime \, e_p^\prime \ ,
\quad p=1,\ldots,k_1-1,k_1+1,\ldots,k_1+k_2 
\label{der5}
\end{equation}
for the $(k_1-1+k_2)$ radial Gelfand--Tzetlin
coordinates $s_p^\prime$ and the corresponding vectors $e_p^\prime$ as
eigenvalues and eigenvectors of the matrix $(1_{k_12k_2}-u_{1}u_{1}^\dagger)
\, s \, (1_{k_12k_2}-u_{1}u_{1}^\dagger)$
which has the generalized rank $k_1-1+k_2$.  Due to
$u_{1}^\dagger e_p^\prime = 0$, we find
\begin{equation}
(1_{k_12k_2}-u_1u_1^\dagger) \, s \, e_p^\prime \ = \
                            s_p^\prime \, e_p^\prime \ ,
\quad p=1,\ldots,k_1-1,k_1+1,\ldots,k_2 \ .
\label{der6}
\end{equation}
As in Ref.~\cite{GGT}, the eigenvalues $s_p^\prime$ are calculated
from the characteristic function
\begin{eqnarray}
z(s_p^\prime) 
  &=& \detg\left((1_{k_12k_2}-u_1u_1^\dagger)s-s_p^\prime\right)
                               \nonumber\\
  &=& -s_p^\prime \, \detg\left(s-s_p^\prime\right) \,
         u_{1}^\dagger\frac{1_{k_12k_2}}{s-s_p^\prime}u_{1} 
\label{der7}
\end{eqnarray}
which has to be discussed in the limits
\begin{eqnarray}
z(s_p^\prime) \ \longrightarrow \ 
          \cases{ 0       & \ for \  $p=1,\ldots,k_1-1$ \cr
                  \infty  & \ for \  $p=k_1+1,\ldots,k_1+k_2$ } \quad .
\label{der8}
\end{eqnarray}
Thus, together with the normalization  $u_{1}^\dagger u_{1}=1$,
these are $k_1+k_2$ equations for the elements of $u_{1}$.

The two parts of the integral~(\ref{der4}) have to be expressed in
terms of the radial Gelfand--Tzetlin coordinates $s_p^\prime$. In a
calculation fully analogous to the ordinary case, we find
\begin{equation}
\sigma_{11} \ = \ \trg s \, - \, \trg s^\prime \ .
\label{der9}
\end{equation}
The eigenvalues $t_p, \ p=1,\ldots,k_1-1,k_1+1,\ldots,k_1+k_2$ of
$\widetilde{\sigma}$ obtain from the characteristic function
\begin{eqnarray}
w(t_p) \ = \ \detg\left(\widetilde{\sigma}-t_p\right)
  &=& - \frac{1}{t_p} \, 
        \detg\left((1_{k_12k_2}-u_{1}u_{1}^\dagger)s-t_p\right)\quad . 
\label{der10}
\end{eqnarray}
Comparison with Eq.~(\ref{der7}) shows that the characteristic
functions $w(t_p)$ and $z(s_p^\prime)$ are, apart from the non--zero 
factor $-t_p$, identical. This implies $t_p\equiv s_p^\prime, \  
p=1,\ldots,k_1-1,k_1+1,\ldots,k_1+k_2$. Thus, by introducing the 
square matrix $\widetilde{u}$ which diagonalizes $\widetilde{\sigma}$, 
we may write
\begin{equation}
\widetilde{\sigma} \ = \ b^\dagger s b 
              \ = \ \widetilde{u}^\dagger s^\prime \widetilde{u} \ .
\label{der11}
\end{equation}
By construction, $\widetilde{u}$ must be in the group ${\it
  UOSp}(k_1-1/2k_2)$, because $\sigma$ and $\widetilde{\sigma}$ share
the same symmetries.

These intermediate results allow us to transform Eq.~(\ref{der4}) into
\begin{eqnarray}
\Phi_{k_12k_2}(s,r) &=& \int d\mu(s^\prime,s) \, 
                        \exp(i(\trg s - \trg s^\prime)r_{11}) \,
                        \int d\mu(b) \, 
              \exp(i\trg\widetilde{u}^\dagger s^\prime\widetilde{u} 
                                              \widetilde{r})\cr
                  && 
\label{der12}
\end{eqnarray}
where $d\mu(s^\prime,s)$ is, apart from phase angles, the invariant
measure $d\mu(u_1)$, expressed in the radial Gelfand--Tzetlin
coordinates $s^\prime$. To do the integration over $b$, we view, for
the moment, the vector $u_1$ as fixed and observe that the measure
$d\mu(b)$ is the invariant measure of the group ${\it
  UOSp}(k_1-1/2k_2)$ under the constraint that $b$ is locally
orthogonal to $u_1$. The matrix $\widetilde{u}\in{\it
  UOSp}(k_1-1/2k_2)$ is constructed from $b$ under the same
constraint. Thus, since $b$ and $\widetilde{u}$ cover the same
manifold, the integral over $b$ in Eq.~(\ref{der12}) must yield the
supermatrix Bessel function
$\Phi_{(k_1-1)2k_2}(s^\prime,\widetilde{r})$ and we arrive at the
supersymmetric recursion formula~(\ref{susyrec}).  In the last step,
we used a line of arguing slightly different from the derivation in
ordinary space. In this way we avoided a discussion related to the
ill--defined supergroup volume. The invariance of the measure is the
crucial property we need for the proof and this holds both in
superspace and in ordinary space.

\subsection{Invariant Measure}
\label{inv}

In order to evaluate the invariant measure, we have to solve the
system of equations~(\ref{der7}) for $|v_p^{(1)}|^2=|u_{p1}|^2$,
$p=1,\ldots,k_1$ and
$|\alpha_p^{(1)}|^2=|u_{(k_1+2p)1}|^2+|u_{(k_1+2p-1)1}|^2$,
$p=1,\ldots,k_2$ in terms of the bosonic eigenvalues $s_{p}^\prime=s_{p1}^\prime$,
$p=1,\ldots,k_1-1$ and the fermionic eigenvalues
$s_{k_1+2p}^\prime=s^\prime_{k_1+2p-1}=is_{p2}^\prime$, $p=1,\ldots,k_2$.
\begin{eqnarray}
1 & = & \sum_{p=1}^{k_1} |v_p^{(1)}|^2 \, 
         + \, \sum_{p=1}^{k_2} |\alpha_p^{(1)}|^2  ,\label{ss8}\\
0 & = & \sum_{q=1}^{k_1}
        \frac{|v_q^{(1)}|^2}{s_{q1}-s_{p1}^{\prime}} + 
        \sum_{q=1}^{k_2} 
        \frac{|\alpha_q^{(1)}|^2}
             {is_{q2}-s_{p1}^{\prime}} 
        , \quad
        p=1,\ldots,k_1-1 \ ,\label{ss9}\\
z_p & = & is_{p2}^{\prime}
          \frac{\prod_{q=1}^{k_1}
          (s_{q1}-is_{p2}^{\prime})}
          {\prod_{q=1}^{k_2}
          (is_{q2}-is_{p2}^{\prime})^2}
          \left(\sum_{q=1}^{k_1}  
          \frac{|v_q^{(1)}|^2}{s_{q1}-is_{p2}^{\prime}} + 
          \sum_{q=1}^{k_2} 
         \frac{|\alpha_q^{(1)}|^2}{is_{q2}-is_{p2}^{\prime}}
         \right) \ ,\cr
    & & \qquad \qquad \qquad \qquad \qquad \qquad \quad 
                         z_p\to\infty, \quad p=1,\ldots,k_2 \ . 
\label{ss10}
\end{eqnarray}
In App.~\ref{appCC1}, we sketch the solution of this system for small
dimensions. Inspired by these solutions one can conjecture the general
solutions and verify them by plugging them directly into
Eq.~(\ref{ss8}) to~(\ref{ss10}) , one finds
\begin{eqnarray}
  |v_p^{(1)}|^2 & = & \frac{\prod_{q=1}^{k_1-1}
    (s_{p1}-s_{q1}^{\prime}) \prod_{q=1}^{k_2}(s_{p1}-is_{q2})^2}
  {\prod_{q=1}^{k_2}(s_{p1}-is_{q2}^{\prime})^2 \prod_{q=1,q\neq
      p}^{k_1}(s_{p1}-s_{q1})}\quad, \quad p=1,\ldots,k_1 \ ,\cr
  |\alpha_p^{(1)}|^2 & = & 2\,(is_{p2}^{\prime}-is_{p2})
  \frac{\prod_{q=1}^{k_1-1}(is_{p2}-s_{q1}^{\prime}) \prod_{q=1,q\neq
      p}^{k_2}(is_{p2}-is_{q2})^2} {\prod_{q=1,q\neq
      p}^{k_2}(is_{p2}-is_{q2}^{\prime})^2 \prod_{q=1}^{k_1}
    (is_{p2}-s_{q1})}\, ,\cr && \qquad \qquad \qquad\qquad\qquad
  \qquad \qquad\qquad \qquad p=1,\ldots,k_2 \quad.
\label{ss5}
\end{eqnarray}
These expressions are reminiscent of the ones derived in
Ref.~\cite{GGT} for unitary matrices. However, importantly, all
products in (\ref{ss5}) involving fermionic eigenvalues are squared.
This reflects the degeneracy of $s$ in the fermion-fermion block. We
have introduced a new anticommuting variables $\xi_p^\prime,
\xi_p^{\prime*}$ with $|\xi_p^\prime|^2=is_{p2}^\prime-is_{p2}$
according to definition~(\ref{ss6}).  

{}From this point on, the invariant measure can be calculated in the
same way as for the angular Gelfand--Tzetlin coordinates, see
Ref.~\cite{GGT} for details. 
The result is summarized in Eqs.~(\ref{ss7}).

\section{The Function $\Phi_{22}(\s,\r)$}
\label{subs322}

We use the recursion formula (\ref{susyrec}) to calculate the
supermatrix Bessel function for ${\it UOSp}(2/2)$. To avoid the
imaginary unit in the exponent, we study $\Phi_{22}(-is,r)$. The
recursion formula reads
\begin{equation}
\Phi_{22}(-is,r) \ = \ \widehat{G}_{22}
               \int d\mu(s^\prime,s)\,\exp((\trg s -\trg s^\prime)r_{11})\;  
                \Phi_{12}(-is^\prime,\widetilde{r}) \ .
\label{lde1}
\end{equation}
The function $\Phi_{12}(-is^\prime,\widetilde{r})$ is easily found to be 
\begin{equation}
\Phi_{12}(-is^\prime,\widetilde{r}) \ = \ \widehat{G}_{12}
          \bigg(1-2(r_{21}-ir_{12})(s_{11}^\prime-is_{12}^\prime)\bigg)\ 
            \exp(2r_{12}s_{12}) \ .
\label{lde2}
\end{equation}
The measure of the coset ${\it UOSp}(2/2)/{\it UOSp}(1/2)$ 
is according to formula (\ref{ss7}) given by
\begin{equation}
d\mu(s^\prime,s)=  \frac{(is_{12}-s_{11}^\prime)
                        \prod_{n=1}^2(is_{12}^\prime-s_{n1})}
                       { \sqrt{-\prod_{n=1}^2(s_{11}^\prime-s_{n1})}
                        \left(is_{12}^\prime-s_{11}^\prime\right)^2}
                  ds_{11}^\prime d\xi_1^{\prime*}d\xi_1^\prime \ .
\label{lde3}
\end{equation}
We do the Grassmann integration and find
\begin{eqnarray}
\lefteqn{\Phi_{22}(-is,r) =  \widehat{G}_{22}\,
               \exp\left(r_{11}(s_{11}+s_{21})-2is_{12}ir_{12}\right)}\nonumber\\ 
   & &\int_{s_{11}}^{s_{21}} \mu_B(s^\prime,s)
      \prod_{q=1}^2(is_{12}-s_{q1})
                     \biggl( 4\prod_{j=1}^2(ir_{12}-r_{j1}) \biggr.
                            \nonumber\\
   & & \biggl. -2(ir_{12}-r_{11})\sum_{q=1}^2\frac{1}{is_{12}-s_{q1}}+
                   2 M_{11}(s_1^\prime,s_{1})\biggr)
       \exp\left(s_{11}^\prime(r_{21}-r_{11})\right)ds_{11}^\prime \
       ,\nonumber\\
   & &
\label{lde4}
\end{eqnarray}
where we have introduced the operator
\begin{eqnarray}
\lefteqn{M_{mj}(s_1^\prime,s_1) = \frac{1}{(is_{m2}-s_{j1}^\prime)}}\cr
                &&  \qquad      \left(
                        \frac{1}{2}\sum_{n=1}^{k_1}\frac{1}{is_{m2}-s_{n1}}
                        - \frac{1}{is_{m2}-s_{j1}^\prime}
                        -\sum_{n=1 \atop n\neq j}^{k_1}
                        \frac{1}{s_{j1}^\prime-s_{n1}^\prime}
                        -\frac{\partial}{\partial s_{j1}^\prime}
                        \right)\ .
\label{lde5}
\end{eqnarray}
For later purposes, we introduced general indices $m$ and $j$.
Obviously, the Grassmann integration yielded eigenvalues in the
denominator. This is somewhat surprising because of the following
observation: we can always parametrize the group element $u\in {\it
  UOSp}(2/2)$ in a non--canonical coset parametrization in the spirit
of an Euler parametrization in ordinary space.  Inserting this
parametrization into the defining equation of the supermatrix Bessel
function (\ref{321}) one can expand the trace in all Grassmann
variables. The expansion coefficients are polynomials in the commuting
integration variables and -- more important -- in the matrix elements
of $s$ and $r$. The invariant measure can be expanded in the Grassmann
variables as well. It does not depend on $r$ and $s$.  Although this
procedure becomes rapidly out of hand even for small groups, it is
clear that the outcome of this expansion will be polynomial in the
eigenvalues of $s$ and $r$.  In other words: eigenvalues can only
appear in the denominator by an integration over commuting variables
and never by a Grassmann integration. Therefore, before performing any
integral over commuting variables, there must exist a form of
$\Phi_{22}(-is,r)$, which is polynomial in the eigenvalues of $s$ and
$r$.
 
To remove the denominators and to obtain such a polynomial expression,
we use the following result: Let $f(s_1^\prime)$ be an analytic,
symmetric function in $s_{1i}^\prime, i=1,\ldots k_1$. Furthermore,
define the operator
\begin{equation}
L_m(s)=\sum_{j=1}^{k_1}\frac{1}{is_{m2}-s_{j1}}
    \frac{\partial}{\partial s_{j1}}\qquad.
\label{lde6}
\end{equation}
Then the action of the operator on the integral over the bosonic
part of the measure is given by
\begin{eqnarray}
\lefteqn{L_m(s) \int_{s_{11}}^{s_{21}}\ldots \int_{s_{(k_1-1)1}}^{s_{k_11}}
        \mu_B(s^\prime,s)\;f(s_1^\prime)\;d[s_1^\prime] =}\cr 
 &&   -\int_{s_{11}}^{s_{21}}\ldots \int_{s_{(k_1-1)1}}^{s_{k_11}}
         \mu_B(s^\prime,s)
         \sum_{j=1}^{k_1-1}M_{mj}(s_1^\prime,s_1)\;f(s_1^\prime)\;d[s_1^\prime]
         \quad .
\label{LEM1}
\end{eqnarray}
This formula is derived in App.~\ref{derilem1}.

We now set $f(s_{1}^\prime)=\exp(-s_{11}^\prime(r_{21}-r_{11}))$ and
insert Eq.~(\ref{LEM1}) into Eq.~(\ref{lde4}), we arrive at
\begin{eqnarray}
\lefteqn{\Phi_{22}(-is,r)  = \widehat{G}_{22} \exp(-2is_{12}ir_{12})
          \left( 4\prod_{j=1}^2(ir_{12}-r_{j1})(is_{12}-s_{j1})-\right.}\cr
               &   &\left.2\sum_{q=1}^2 (is_{12}-s_{q1})
                    \left(ir_{12}-r_{21}-r_{11}-
                    \frac{\partial}{\partial s_{q1}}\right)\right)
                    \Phi_{2}^{(1)}(-is_1, r_1)\ ,
\label{lde10}
\end{eqnarray} 
where $\Phi_{2}^{(1)}(s_1,r_1)$ is the matrix Bessel function of the
orthogonal group ${\it O}(2)$ in ordinary space as defined in
Ref.~\cite{GK1}.  Although this can already be taken as the result, we
underline the symmetry between $s$ and $r$ by using the explicit
form~(3.21) of Ref.~\cite{GK1} for $\Phi_{2}(s_1, r_1)$
\begin{eqnarray}
\Phi_{22}(-is,r) &=& \widehat{G}_{22} \exp\left(\trg rs -
                            \frac{z}{2}\right)\cr
                 & & \ \biggl(
                           4\prod_{j=1}^2(ir_{21}-r_{1j})(is_{21}-s_{1j})
                          -\sum_{q=1}^2 (is_{21}-s_{1q})       
                           \sum_{p=1}^2 (ir_{21}-r_{1p}) \biggr. \cr
                 & & \ \qquad\qquad\qquad \biggl.
                            -z\frac{d}{dz}\biggr)2\pi I_0(z/2) \ ,
\label{lde11}
\end{eqnarray}
where we have introduced $z=(s_{11}-s_{12})(r_{11}-r_{12})$ and the
modified Bessel function $I_0$ as defined in Ref.~\cite{AS}.

The result~(\ref{LEM1}) was crucial in the derivation of
$\Phi_{22}(-is,r)$. By means of this formula, the denominator problem
was overcome in one step. Because of its importance, we want to gain
more insight into this problem: In App.~\ref{altder}, we rederive
$\Phi_{22}(-is,r)$ in two other ways. It is clear that the
methods of App.~\ref{altder} can not be used for higher dimensions
$k_1$ and $2k_2$, but it will help to understand the mechanisms needed
when working with radial Gelfand--Tzetlin coordinates.

\section{The Series of Functions $\Phi_{\k_14}(\s,\r)$}
\label{subs323}

We calculate iteratively the four supermatrix Bessel functions
$\Phi_{k_14}(s,r)$ for $k_1=1,2,3,4$. We do this in 
Secs.~\ref{ser1} to~\ref{ser4}, respectively.

\subsection{First Level $k_1=1$}
\label{ser1}

According to the recursion formulae~(\ref{susyrec}) and~(\ref{ss6aa}),
the starting point is the matrix Bessel function for the unitary
symplectic group $\Phi_2^{(4)}(s_2,r_2)$, which was already
calculated in Ref.~\cite{GK1}. Up to a normalization, we have
\begin{eqnarray}
\Phi^{(4)}_{2}(i2s_2,r_2)&=& \sum_{\omega\in S_2}
              \left(
              \frac{1}{\Delta_2^2\left(is_2\right)
                       \Delta_2^2\left(\omega(ir_2)\right)}-
              \frac{1}{\Delta_2^3\left(is_2\right)
                       \Delta_2^3\left(\omega(ir_2)\right)} 
              \right) \nonumber\\
           & &\qquad\qquad\qquad\qquad
              \exp(-\tr is_2\omega(ir_2)) \ ,
\label{usp3}
\end{eqnarray}
Since the subgroup ${\it O}(1)$ of ${\it UOSp}(1/4)$ is trivial, no
commuting integral has to be performed to derive $\Phi_{14}(-is,r)$.
Inserting the measure~(\ref{ss7}) into the recursion formula and
performing the Grassmann integrations yields straightforwardly
\begin{eqnarray}
\Phi_{14}(-is,r)&=&\widehat{G}_{14} \exp\left(\trg rs \right)
                 \left(\frac{1}{\Delta_2^2(ir_2)\Delta_2^2(is_2)}+
                 \frac{1}{\Delta_2^3(ir_2)\Delta_2^3(is_2)}\right)\cr
              & &\bigg(2(is_{21}-s_{11})(ir_{21}-r_{11})-1\bigg)
                 \bigg(2(is_{22}-s_{11})(ir_{22}-r_{11})-1\bigg)\cr
              & & \qquad\qquad 
                  - \widehat{G}_{14}\frac{\exp\left(\trg(rs)\right)}  
                               {\Delta_2^3(ir_2)\Delta_2^3(is_2)}
                 \, + \,
                \left( ir_{12} \longleftrightarrow ir_{22}\right) \ .
\label{p41}
\end{eqnarray} 
The exchange term $\left( ir_{12} \longleftrightarrow ir_{22}\right)$\ 
accounts for the permutation group $S_2$ in Eq.~(\ref{usp3}).
Anticipating that the structure of $\Phi_{14}(-is,r)$ will,
remarkably, survive on all levels up to $\Phi_{44}(-is,r)$, we state
that $\Phi_{14}(-is,r)$ essentially consists of two parts. A
comparison with Eqs.~(\ref{lde2}), and~(\ref{usp3}) shows, that the
first part of $\Phi_{14}(-is,r)$ is a product of an exponential with
three other terms. The first one,
\begin{equation}
\left(\frac{1}{\Delta_2^2(ir_2)\Delta_2^2(is_2)}+
                 \frac{1}{\Delta_2^3(ir_2)\Delta_2^3(is_2)}\right)\quad ,
\label{p41aa}
\end{equation}
stems from the integral over the ${\it USp}(4)$ subgroup. The other
two terms can be identified with the supermatrix Bessel functions
\begin{equation} 
\Phi_{12}(-is,r) \quad {\rm with} 
                      \quad  s=\diag(s_{11},is_{12},is_{12}) , \
                             r=\diag(r_{11},ir_{12},ir_{12})
\label{p41bb}
\end{equation}
and
\begin{equation} 
\Phi_{12}(-is,r) \quad {\rm with} 
                      \quad  s=\diag(s_{11},is_{22},is_{22}) , \
                             r=\diag(r_{11},ir_{22},ir_{22}) \ .
\label{p41c}
\end{equation}
The second part can be considered as a correction term, which destroys
the product structure of $\Phi_{14}(-is,r)$.  We may identify the
different parts of the product with the integrations over the
corresponding subsets of the group. Thus, $\Phi_{2}^{(4)}(-is_2,r_2)$
arises from the integration over the ${\it USp}(4)$ subgroup, the
${\it O}(1)$ integration yields unity and the other two factors come
from the integration over the coset 
${\it UOSp}(1/4)/({\it USp}(4)\otimes{\it O}(1))$.

\subsection{Second Level $k_1=2$}
\label{ser2}
  
We now have to do one integration over a commuting variable.  After
the Grassmann integration, we are left with a considerable amount of
terms.  To arrange them in a convenient way we, introduce the
following notation for the product of two operators $D_1(s) D_2(s)$
acting on a function $f(s)$, we define
\begin{equation}
\left[D_1^\to(s)\, D_2(s)\right] f(s)= D_1(s)\, D_2(s)\, f(s) 
                 - \left(D_1(s)\, D_2(s)\right)\,f(s)\quad .
\label{p41b}
\end{equation}
This means, an operator with an arrow only acts on the terms outside
the squared bracket. With this notation we can write 
\begin{eqnarray}
\Phi_{24}(-is,r)&=&\widehat{G}_{24} 
                  \exp\left(\tr(r_2s_2)+r_{21}(s_{11}+s_{21})\right) 
                                        \nonumber\\
                & & \qquad\qquad
                   \int_{s_{11}}^{s_{21}}d\mu_B(s_1^\prime,s_1)\left[
                   \prod_{i=1}^2\prod_{j=1}^2(is_{i2}-s_{j1})\right.
                                        \nonumber\\
                & & \qquad\qquad\qquad
                   \left(\frac{1}{\Delta_2^2(ir_2)\Delta_2^2(is_2)}+
                   \frac{1}{\Delta_2^3(ir_2)\Delta_2^3(is_2)}\right)
                                        \nonumber\\
                & & \qquad\qquad\left(4\prod_{i=1}^2(ir_{12}-r_{i1})
                   +2\sum_{k=1}^2\frac{r_{21}-ir_{12}}{is_{12}-s_{k1}}
                   +2 M_{11}^\rightarrow(s_1^\prime,s_1)\right)
                                        \nonumber\\
              &  & \qquad\qquad\left(4\prod_{i=1}^2(ir_{22}-r_{i1})
                   +2\sum_{k=1}^2\frac{r_{21}-ir_{22}}{is_{22}-s_{k1}}
                   +2 M_{21}(s_1^\prime,s_1)\right)+
                                        \nonumber\\            
              &  & \left(\frac{1}{\Delta_2^2(ir_2)\Delta_3^2(is_2)}+
                   \frac{1}{\Delta_2^3(ir_2)\Delta_2^4(is_2)}\right)
                                        \nonumber\\ 
              &  & \left(\frac{4}{is_{12}-s_{11}^\prime}M_{21}(s_1^\prime,s_1)-
                   \frac{2}{is_{22}-s_{11}^\prime}
                   M_{11}(s_1^\prime,s_1)\right)
                                        \nonumber\\
              &  & +\frac{2}{\Delta_2^3(ir_2)\Delta_2^4(is_2)}
                   \left( 2 \tr r_1 - \tr ir_2 
                   +\sum_{i=1}^2\frac{1}{is_{22}-s_{i1}}\right)
                   M_{11}(s_1^\prime,s_1)
                                        \nonumber\\
              &  & -\frac{2}{\Delta_2^3(ir_2)\Delta_2^4(is_2)}
                   \left( 2 \tr r_1 -\tr ir_2
                   +\sum_{i=1}^2\frac{1}{is_{12}-s_{i1}}\right)
                   M_{21}(s_1^\prime,s_1) 
                                        \nonumber\\
              &  & \left.-\frac{4}{\Delta_2^3(ir_2)\Delta_2^3(is_2)}
                   \sum_{k=1}^2\prod_{j=1}^2
                   \frac{r_{21}-ir_{j2}}{s_{k1}-is_{j2}}\right]
                   \;\exp\left(s_{11}^\prime(r_{11}-r_{21})\right)
                                        \nonumber\\
              &  & \qquad\qquad\qquad
                  + \left( ir_{12} \longleftrightarrow ir_{22}\right)\quad .
\label{p42}
\end{eqnarray}
As in Sec.~\ref{subs322}, a denominator problem occurs. It becomes
obvious in the product
$M_{11}^\rightarrow(s_1^\prime,s_1)M_{21}(s_1^\prime,s_1)$.  Thus, we
expect an identity similar to formula~(\ref{LEM1}). This identity
should map a product of operators $L_1(s)L_2(s)$ acting on the
integral onto a product of operators
$M_{11}(s_1^\prime,s_1)M_{21}(s_1^\prime,s_1)$ acting under the
integral. Neither the outer operators, $L_m(s)$, nor the inner ones,
$M_{mj}(s)$, commute. Hence, the desired identity must be a
non--trivial one. It is given by the following result.

We have the same conditions as in formula~(\ref{LEM1}), furthermore we
define
\begin{equation}
\left[L_m^\to(s)L_l(s)\right]=
\sum_{n=1}^{k_1}\sum_{q=1}^{k_1}\frac{1}{(is_{m2}-s_{n1})(is_{l2}-s_{q1})}
\frac{\partial^2}{\partial s_{n1}\partial s_{q1}}\quad .
\label{p43}
\end{equation} 
Then the following formula holds
\begin{eqnarray}
\lefteqn{\left[L_m^\to(s) L_l(s)\right]\int_{s_{11}}^{s_{21}}\!\ldots\! 
        \int_{s_{(k_1-1)1}}^{s_{k_11}}
        \mu_B(s^\prime,s)\, d[s_1^\prime]\, f(s_1^\prime) =}\cr 
 &&     \int_{s_{11}}^{s_{21}}\ldots \int_{s_{(k_1-1)1}}^{s_{k_11}}
         \mu_B(s^\prime,s)\left[
         \sum_{j=1}^{k_1-1}\sum_{k=1}^{k_1-1}M_{mj}^\to(s_1^\prime,s_1)
         M_{lk}(s_1^\prime,s_1)-\right.\cr
 &&     \frac{1}{is_{l2}-is_{m2}}\sum_{j=1}^{k_1-1}\left(
        \frac{1}{is_{m2}-s_{j1}^\prime}M_{lj}(s_1^\prime,s_1)-
        \frac{1}{is_{l2}-s_{j1}^\prime}M_{mj}(s_1^\prime,s_1)\right)\cr
 &&     \left.-\frac{1}{2}\sum_{k\neq j}^{k_1-1}
        \frac{1}{(is_{m2}-s_{k1}^\prime)(is_{m2}-s_{j1}^\prime)
        (is_{l2}-s_{k1}^\prime)(is_{l2}-s_{j1}^\prime)}\right]\,
        f(s_1^\prime)\,d[s_1^\prime] \ . \cr
 &&
\label{lem3}
\end{eqnarray}
The derivation is along the same lines as the one for
formula~(\ref{LEM1}), it also involves formula~(\ref{LEM1}). With the
identities~(\ref{lem3}) and~(\ref{LEM1}) the denominator problem is
again solved in one step. After some further manipulations we arrive
at
\begin{eqnarray}
\Phi_{24}(-is,r)&=& 2\pi\widehat{G}_{24} 
               \exp\left(\trg rs -\frac{z}{2}\right) \cr
                & & \qquad\qquad
                 \left(\frac{1}{\Delta_2^2(ir_2)\Delta_2^2(is_2)}+
                   \frac{1}{\Delta_2^3(ir_2)\Delta_2^3(is_2)}\right)
                                                 \cr
              &&\left[\biggl(4\prod_{i=1}^2(r_{i1}-ir_{12})(s_{i1}-is_{12})
                                \biggr.\right. \cr
              && \qquad\qquad\qquad \biggl.
                -\sum_{i=1 \atop j=1}^2(s_{j1}-is_{12})(r_{i1}-ir_{12})
                -z\frac{\partial^\to}{\partial z}\biggr) \cr
              & & \qquad 
                    \biggl(4\prod_{i=1}^2(r_{i1}-ir_{22})(s_{i1}-is_{22})
                         \biggr. \cr
              & &\qquad\qquad \left.\biggl.
                 -\sum_{i=1 \atop j=1}^2(s_{j1}-is_{22})(r_{i1}-ir_{22})
                 -z\frac{\partial}{\partial z}\biggr)\right]I_0(z/2)
                                 \cr
              &&-2\pi\widehat{G}_{24}\exp\left(\trg rs -\frac{z}{2}\right)
                                 \cr
              && \qquad
                 \frac{2}{\Delta_2^3(ir_2)\Delta_2^3(is_2)}
                 \sum_{i=1 \atop k=1}^2\prod_{j=1}^2(s_{i1}-is_{j2})
                  (r_{k1}-ir_{j2})I_0(z/2)\cr
              &&-2\pi\widehat{G}_{24}\exp\left(\trg rs -\frac{z}{2}\right)
                                 \cr
              && \qquad
                 \frac{1}{\Delta_2^3(ir_2)\Delta_2^3(is_2)}
                 \left((\trg s) (\trg r) -1\right) z
                 \frac{\partial}{\partial z} I_0(z/2)\cr 
              & & \qquad\qquad \qquad\qquad  \qquad\qquad   
            + \ \left( ir_{12} \longleftrightarrow
              ir_{22}\right) \ .
\label{p45}
\end{eqnarray}
As in Sec.~\ref{subs322}, we used the composite variable
$z=(s_{11}-s_{12})(r_{11}-r_{12})$.  A comparison with
Eqs.~(\ref{lde10}) and~(\ref{p41}) shows the similarity in the
structures of $\Phi_{24}(-is,r)$ and $\Phi_{14}(-is,r)$. The former
also decomposes into two parts. The first part is a product, whose
factors can be assigned to the integrations over the different
submanifolds of the group in the same way as in the case of
$\Phi_{14}(-is,r)$. The other one can be interpreted as a correction
term due to the non--commutativity of the operators $L_m$ in
formula~(\ref{lem3}).

\subsection{Third Level $k_1=3$}
\label{ser3}

This structure of $\Phi_{k_14}(-is,r)$ emerging in the previous
calculations is likely to be also present for arbitrary $k_1$.
However, for $k_1 > 2$, we have so far not been able to treat the
general case.  Fortunately, in important physics applications, one
matrix argument of the supermatrix Bessel function has an additional
twofold degeneracy in the boson--boson block.  In this case, it is
possible to carry on the recursion up to $\Phi_{44}(-is,r)$ by
extending the techniques developed for $k_1=1$ and $k_2=2$.  Thus,
from now on, we restrict ourselves to this case.

At first sight, one might hope to achieve some simplification by
applying the projection procedure onto the degenerate matrix, because
this results in a considerable simplification of the invariant
measure. However, it turned out that the integrations are easier if
one does the recursion with the non--degenerate coordinates.  Hence,
we use the measure as it stands in Eq.~(\ref{ss7}).  We consider
$\Phi_{34}(-is,r)$ in the case that
\begin{equation}
r_1 = 
      \diag(r_{11},r_{21},r_{21})\quad .
\label{p46}
\end{equation}
Having performed the Grassmann integral, one can arrange the terms in
a way similar to Eq.~(\ref{p42}). The complete expression and further
details are given in App.~\ref{appH3}. We then can use
formula~(\ref{lem3}) and find after some further algebra
\begin{eqnarray}
\Phi_{34}(-is,r)&=& 4\;\widehat{G}_{34}\exp\left(\tr r_2s_2\right)
                          (r_{21}-ir_{12})(r_{21}-ir_{22})\cr
              & &
                 \Biggl[\left(\frac{1}{\Delta_2^2(ir_2)\Delta_2^2(is_2)}+
                 \frac{1}{\Delta_2^3(ir_2)\Delta_2^3(is_2)}\right)\Biggr.\cr
              & &\left(4\prod_{k=1}^2(r_{k1}-ir_{12})
                 \prod_{k=1}^3(s_{k1}-is_{12})\right.\cr
              && \qquad \left.
                 +2 \sum_{k=1}^3\prod_{j\neq k}^3(s_{j1}-is_{12})
                 \left(r_{11}+r_{21}-ir_{12}
                 -\frac{\partial^\to}{\partial s_{k1}}\right)\right)\cr
              & &\left(4\prod_{k=1}^2(r_{k1}-ir_{22})
                 \prod_{k=1}^3(s_{k1}-is_{22})\right.\cr
              && \qquad \left.
                 +2 \sum_{k=1}^3\prod_{j\neq k}^3(s_{j1}-is_{22})
                 \left(r_{11}+r_{21}-ir_{22}
                 -\frac{\partial}{\partial s_{k1}}\right)\right)\cr
              & &\qquad\qquad\qquad-\frac{4}
                 {\Delta_2^3(ir_2)\Delta_2^3(is_2)}
                 \sum_{i=1}^3\prod_{j=1 \atop j\neq i}^3
                 (s_{j1}-is_{12})(s_{j1}-is_{22})\cr
              & &\biggl(r_{11}^2+r_{21}^2+r_{11}r_{21}-
                 (r_{11}+r_{21})(ir_{12}+ir_{22})+\biggr.\cr
              && \qquad\qquad \biggl.ir_{12}ir_{22}-
                 (r_{11}+r_{21}-ir_{12}-ir_{22})
                 \frac{\partial}{\partial s_{i1}}\biggr)\cr 
              & &-\frac{2}{\Delta_2^3(ir_2)\Delta_2^4(is_2)}
                 \sum_{i,k}^3
                 \prod_{j=1 \atop j\neq i}^3
                 \prod_{l=1 \atop l\neq k}^3(s_{j1}-is_{12})(s_{l1}-is_{22})\cr
              & &\qquad\qquad\qquad\Biggl.
                 \left(\frac{\partial}{\partial s_{i1}}-
                 \frac{\partial}{\partial s_{k1}}\right)\Biggr] \,
                 \Phi_3^{(1)}(-is_1,r_1)  \cr
              && \qquad\qquad\qquad  \, + \, 
                 \left( ir_{12} \longleftrightarrow ir_{22}\right) \ ,
\label{p49}
\end{eqnarray}
where $\Phi_3^{(1)}(s_1,r_1)$ is the matrix Bessel function of the
orthogonal group ${\it O}(3)$. We notice, that the structure of
$\Phi_{14}(-is,r)$ and $\Phi_{24}(-is,r)$ reappears in
$\Phi_{34}(-is,r)$.

\subsection{Fourth Level $k_1=4$}
\label{ser4}

In the calculation of $\Phi_{44}(-is,r)$, we again consider the case
that the matrix $r$ is degenerate,
\begin{equation}
r_1 = 
      \diag(r_{11},r_{11},r_{21},r_{21})\quad .
\label{p410}
\end{equation}
The main problem is to find a convenient representation for the matrix
Bessel function $\Phi_3^{(1)}(-is_1^\prime,\widetilde{r}_1)$ appearing
on the third level in Eq.~(\ref{p49}).  It turns out that the
representation derived in App.~B of Ref.~\cite{GK1} is very well
suited to our purpose. Due to the degeneracy in $\widetilde{r}_1$, the
original threefold integral can be reduced to an integral over just
one single variable
\begin{eqnarray}
\Phi_3^{(1)}(-is_1^\prime,\widetilde{r}_1) &=& 
                     \frac{\exp\left(r_{21}\tr s_1^\prime\right)}
                          {\sqrt{r_{11}-r_{21}}}
                      \int_{-\infty}^{+\infty} dt
              \frac{\exp\left(i(r_{11}-r_{21})t\right)} 
                   {\prod_{i=1}^3\sqrt{s_{i1}^\prime-it}} \ .
\label{SS1}
\end{eqnarray}
Here, we again neglected the normalization because we want to fix it
afterwards as explained above. Similarly, $\Phi_4^{(1)}(-is_1,r_1)$ can
be written as a double integral,
\begin{eqnarray}
\lefteqn{\Phi_4^{(1)}(-is_1,r_1) = 
           \frac{\exp\left(r_{21} \tr s_1 \right)}
                  {(r_{11}-r_{21})^2}} \cr 
     & & \qquad
   \int_{-\infty}^{+\infty} dt_1 \int_{-\infty}^{+\infty} dt_2
           |t_1-t_2|
          \frac{\exp\left(i(r_{11}-r_{21})(t_1+t_2)\right)} 
               {\prod_{i=1}^4\prod_{n=1}^2\sqrt{s_{i1}-it_n}} \ .
\label{H9}
\end{eqnarray}
Singularities have to be taken care of appropriately.  After inserting
Eq.~(\ref{p45}) into the recursion formula and performing the
Grassmann integration, one can arrange the terms in a similar way as
in the case of $\Phi_{34}(-is,r)$.  At this point, we notice that
formulae~(\ref{LEM1}) and~(\ref{lem3}) need to be supplemented by
further identities. We state the most important one in the following.

The same conditions as for formula~(\ref{LEM1}) apply. Moreover, we
define the operator
\begin{equation}
\widetilde{L}_m(s)=
\sum_{q=1}^{k_1}\frac{1}{is_{m2}-s_{q1}}\frac{\partial^2}{\partial s_{q1}^2}+
\frac{1}{2}\;\sum_{q\neq n}\frac{1}{(is_{m2}-s_{q1})(s_{q1}-s_{n1})}\left(
\frac{\partial}{\partial s_{q1}}-\frac{\partial}{\partial
  s_{n1}}\right) \ .
\label{p411}
\end{equation}
Then we have
\begin{eqnarray}
\lefteqn{\widetilde{L}_m(s)\int_{s_{11}}^{s_{11}}\!\ldots\! 
        \int_{s_{(k_1-1)1}}^{s_{k_11}}
        \mu_B(s^\prime,s) d[s_1^\prime] f(s_1^\prime) =}\cr 
 &&     \int_{s_{11}}^{s_{21}}\ldots \int_{s_{(k_1-1)1}}^{s_{k_11}}
        \mu_B(s^\prime,s)\left[\sum_{j=1}M_{mj}^\to(s_1^\prime,s_1)
        \frac{\partial}{\partial s_{j1}^\prime}\right]\;f(s_1^\prime)
        \,d[s_1^\prime] \ .
\label{lem3b}
\end{eqnarray}
Again, the proof is along the same lines as the proof of
formula~(\ref{LEM1}) and the proof of formula (5.9) in Ref.~\cite{GK1}.

Thus, there is a family of rules to transform operators symmetric in $s_{i1}$
$L_m(s), \widetilde{L}_m(s)$ acting onto an integral into an operator acting
under the integral. We need one more such transformation rule which
tells us how the product $\left[L_m(s)^\to\widetilde{L}_l(s)\right]$
transforms into operators acting under the integral. This formula and
further details are given in App.~\ref{appH4}.  Collecting everything,
we finally arrive at
\begin{eqnarray}
\Phi_{44}(-is,r)&=&4\;\widehat{G}_{44}\ \exp\left(-\tr(r_2s_2)\right)
               \prod_{i,j}(r_{1i}-ir_{2j})\cr
             &&\Biggl[      
               \left(\frac{1}{\Delta_2^2(ir_2)\Delta_2^2(is_2)}+
               \frac{1}{\Delta_2^3(ir_2)\Delta_2^3(is_2)}\right)\Biggr.
                                          \cr
             &&\biggl(8\prod_{i=1}^2(r_{i1}-ir_{12})
                       \prod_{j=1}^4(s_{j1}-is_{12})\biggr.\cr
             &&\qquad\biggl.
                   +4\sum_{i=1}^4\prod_{j\neq i}^4(s_{j1}-is_{12})
                   \left(r_{11}+r_{21}-ir_{12}
                   -\frac{\partial^\to}{\partial{s_{i1}}}\right)\biggr)\cr
             &&\biggl(8\prod_{i=1}^2(r_{i1}-ir_{22})
                       \prod_{j=1}^4(s_{j1}-is_{22})\biggr.\cr
             &&\qquad\biggl.
                   +4\sum_{i=1}^4\prod_{j\neq i}^4(s_{j1}-is_{22})
                   \left(r_{11}+r_{21}-ir_{22}-
                        \frac{\partial}{\partial s_{i1}}\right)\biggr)\cr
             &&\qquad-\frac{16}{\Delta_2^3(ir_2)\Delta_2^3(is_2)}
                  \sum_{i=1}^4\prod_{j\neq i}^4(s_{j1}-is_{12})(s_{j1}-is_{22})\cr
             &&\qquad\bigl(r_{11}^2+r_{21}^2+r_{11}r_{21}-(ir_{12}+ir_{22})
                   (r_{11}+r_{21})\bigr.\cr
             &&\qquad \bigl. +ir_{12}ir_{22}
                   +\frac{1}{2} \trg r \frac{\partial}{\partial s_{i1}}
                   \bigr)\cr
             &&\qquad -\frac{8}{\Delta_2^3(ir_2)\Delta_2^4(is_2)}
                   \sum_{i=1 \atop j=1}^4\prod_{l\neq i}^4(s_{l1}-is_{12})
                               \prod_{l\neq j}^4(s_{l1}-is_{22})\cr
             &&\qquad\qquad\Biggl.
                   \left(\frac{\partial}{\partial s_{i1}}-
                   \frac{\partial}{\partial s_{j1}}\right)
                   \Biggr]\Phi_4^{(1)}(-is_1,r_1) \cr
              &  & \qquad\qquad\qquad \, + \, 
            \left( ir_{12} \longleftrightarrow ir_{22}\right) \ .
\label{p413}
\end{eqnarray}
We mention that in the derivation of this result we frequently used properties
of the matrix Bessel functions $\Phi_{3}^{(1)}(s_1,r_1)$ and
$\Phi_{4}^{(1)}(s_1,r_1)$ that only hold for the case that one matrix
has an additional degeneracy.

\section{Asymptotics and Normalization}
\label{asym}

The asymptotic behavior of the supermatrix Bessel functions calculated
in the previous sections is a useful check which also allows us to fix
the normalization constants.  We find from the expressions in
Eqs.~(\ref{lde1}),~(\ref{lde10}) and in
Eqs.~(\ref{p41}),~(\ref{p45}),~(\ref{p49}) and~(\ref{p413})
\begin{eqnarray}
\lim_{s\to \infty \atop r\to \infty}\Phi_{k_12k_2}(-is,r)&=& 2^{k_1k_2}\widehat{G}_{k_12k_2}
                    \frac{\prod_{l=1}^{k_1}\prod_{m=1}^{k_2}
                    \left(s_{l1}-is_{m2}\right)\left(r_{l1}-ir_{m2}\right)}
                    {\Delta_{k_2}^2(is_2)\Delta_{k_2}^2(ir_2)}\cr
              &&
             \det\left[\exp(2s_{i2}r_{j2})\right]_{i,j=1\ldots k_2}\;   
             \lim_{s_{1}\to \infty \atop r_{1} \to \infty}
                 \Phi_{k_1}^{(1)}(-is_1,r_1)\quad .
\label{asym1}
\end{eqnarray}
In the degenerate case, each degenerate eigenvalue contributes
according to its multiplicity.  The asymptotics of the matrix Bessel
functions of the orthogonal group is given by~\cite{HC2,MUI}
\begin{equation}
\lim_{t\to 0}
\Phi_{k_1}^{(1)}(-is_1/t,r_1) \ = \widehat{C}^{(k_1)}
                                  t^{(k_1-1)k_1/4} 
         \frac{\det[\exp(s_{n1}r_{m1}/t)]_{n,m=1,\ldots,k_1}}
              {|\Delta_{k_1}(s_1)\Delta_{k_1}(r_1)|^{1/2}} \ , 
\label{asym2}
\end{equation}
where the constant can be found in Muirhead's book \cite{MUI},
\begin{equation}
\widehat{C}^{(k_1)}=\frac{\Gamma(k_1/2)}{k_1!}\;\pi^{k_1^2/2-k_1/4}\quad .
\label{asym3}
\end{equation}
Thus we find
\begin{eqnarray}
\lefteqn{\lim_{t\to 0} \Phi_{k_12k_2}(-is/t,r)= 
                       2^{k_1k_2}
                       t^{\left((k_1-2k_2)^2+(k_1-2k_2)\right)/4}
                       \widehat{C}^{(k_1)}\widehat{G}_{k_12k_2}}\cr
                 &  & \qquad\qquad
                      \frac{\det[\exp(s_{n1}r_{m1}/t)]_{n,m=1,\ldots,k_1}
                      \det\left[\exp(2s_{i2}r_{j2}/t)\right]_{i,j=1\ldots k_2}}  
                      {\sqrt{\widetilde{B}_{k_1k_2}(s)
                       \widetilde{B}_{k_1k_2}(r)}} \quad .
\label{asym3a}
\end{eqnarray}
for the asymptotic behavior.
 
On the other hand, the supermatrix Bessel function relates to the
kernel of Dyson's Brownian Motion in superspace~\cite{GUH4}.
Due to the normalization of the Gaussian integral,
\begin{equation}
\left(\frac{\pi}{2t}\right)^{-\left((k_1-2k_2)^2+(k_1-2k_2)\right)/4}
           2^{k_2^2-k_2-k_1/2}
\int d[\sigma] \exp\left(-\frac{1}{t}\left(\sigma-\rho\right)\right)
 \ = \ 1
\label{normg}
\end{equation}
the kernel 
\begin{eqnarray}
\Gamma_{k_1k_2}(s,r,t) &=&
\left(\frac{\pi}{2t}\right)^{-\left((k_1-2k_2)^2+(k_1-2k_2)\right)/4}
           2^{k_2^2-k_2-k_1/2} \nonumber\\
 & & \qquad\qquad
\int_{u\in {\it UOSp}(k_1/2k_2)} d\mu(u) 
       \exp\left(-\frac{1}{t}\left(\sigma-\rho\right)\right)
\label{ker}
\end{eqnarray}
is also normalized. Since it is obviously connected with to the
supermatrix Bessel function by
\begin{eqnarray}
\Gamma_{k_1k_2}(s,r,t)&=& 
     \left(\frac{\pi}{2t}\right)^{-\left((k_1-2k_2)^2+(k_1-2k_2)\right)/4}
           2^{k_2^2-k_2-k_1/2}\cr
       &&\qquad\qquad
         \exp \left(-\frac{1}{t}\left(\trg s^2\;+\;\trg r^2\right)\right)
                       \Phi_{k_12k_2}(-is/t,r) \ .
\label{asym4}
\end{eqnarray} 
we can fix the normalization by using the asymptotic behavior
\begin{eqnarray}
\lim_{t\to 0}\Gamma_{k_1k_2}(s,r,t)&=&
     \left(\frac{\pi}{2}\right)^{-\left((k_1-2k_2)^2+(k_1-2k_2)\right)/4}
          \frac{2^{k_2^2-k_2-k_1/2}}{k_1!\;k_2!}\cr
       &&   \frac{\det\left[\delta(s_{i1}-r_{j1})\right]_{i,j=1\ldots k_1}
               \det\left[\delta(s_{i2}-r_{j2})\right]_{i,j=1\ldots k_2}}
              {\sqrt{\widetilde{B}_{k_1k_2}(s)\widetilde{B}_{k_1k_2}(r)}}
              \quad.
\label{asym5}
\end{eqnarray}
of the kernel.  Comparing Eq.~(\ref{asym3a}) with Eq.~(\ref{asym5}),
we find
\begin{equation}
\widehat{G}_{k_12k_2}= \frac{2^{3k_2(k_2-k_1)+k_1^2/4-5k_2/2-k_1/2}}
                        {\pi^{\left((k_1-2k_2)^2+2k_1^2-2k_2\right)/4}
                            k_2!\Gamma(k_1/2)} \ .
\label{asym6}
\end{equation}
We mention that this calculation also shows that the diffusion kernels
of the one--point function and of the two--point function of Dyson's
Brownian motion~\cite{GUH4}, i.e.~the function $\Gamma_{(2k)k}(s,r,t)$
which was denoted by $\Gamma_k(s,r,t)$ in Ref.~\cite{GUH4}, indeed
satisfy the proper initial condition.

\section{Applications}
\label{appl}

Although we focus in this contribution on the mathematical aspects, we
now briefly comment on a particular kind of application.  As the
reader will realize, our results derived in the previous section are,
in some sense, more general than what we need in those applications on
which we focus here. We take this as an indication that explicit
results for even more complex supermatrix Bessel functions can also be
obtained.  The results of the previous sections yield the kernels of
the supersymmetric analogue of Dyson's Brownian Motion for the GOE and
the GSE in the cases $k=1$ and $k=2$.  We do not present the physics
background here. The reader interested in these applications is asked
to consult Refs.~\cite{Haake,MEH,GMGW} for generalities and
Ref.~\cite{GUH4}, in particular Sec.~4.2, for the issue discussed
here. In the present contribution, we use the same notations and
conventions.  We restrict ourselves to the transition towards the GOE
and suppress the index $c$. The corresponding formulae for the
transition towards the GSE are derived accordingly. We treat the one--
and two--point functions in Secs.~\ref{onep} and~\ref{twop},
respectively.

Forrester and Nagao \cite{NF} derived expressions for generalized
one--point functions of Dyson's Brownian motion model with Poissonian
initial conditions. The used an expansions of the Green function in
terms of Jack polynomials. Datta and Kunz \cite{DK} employed a
supersymmetric technique to address the two--level correlation
function of the Poisson GOE transition. They arrive at a finite number
of ordinary and Grassmannian integrals which are still to be
performed.  In our approach, we also arrive at a representation of the
correlation function in terms of a finite number of integrals.
However, since we managed to integrate over almost all angular
integrals in the previous sections, our result contains considerably
less integrals, in particular, no Grassmannian ones. It has a clear
structure due to the fact that, apart from two integrals, all others
are eigenvalue integrals, i.e.~live in the curved eigenvalue space of
Dyson's Brownian motion.  Moreover, since our formulae for the kernel
are valid on all scales, our result is also exact for finite level
number.

\subsection{Level density}
\label{onep}

We use the result~(\ref{lde11}), derived in Sec.~\ref{subs322}, for the
supermatrix Bessel function $\Phi_{22}(-is,r)$.  Using the replacement
$r\longrightarrow (x+J)$ and $s\longrightarrow s/t$ and the
relation~(\ref{asym4}), we obtain the diffusion kernel for the level
density
\begin{eqnarray}
  \lefteqn{\Gamma_1(s,x+J,t) = (2\pi)^{-1/2}\frac{J_1}{2t} }\cr
   & &\exp\left(-\frac{1}{t}(s_{11}-x_1-J_1)^2-
  \frac{1}{t}(s_{21}-x_1-J_1)^2+
  \frac{2}{t}(is_{12}-x_1+J_1)^2\right)\cr 
   & & \qquad\qquad \left(-2
  \frac{J_1}{t}\prod_{j=1}^2(is_{12}-s_{j1})+ \, \sum_{q=1}^2
  (is_{12}-s_{q1})\right)\quad .
\label{op2}
\end{eqnarray}
We take the derivative with respect to the source variable $J_1$
and arrive at the level density
\begin{eqnarray}
  \widehat{R}_1(x_1,t)&= &\frac{1}{(2\pi)^{3/2}t} \int
  \exp\left(-\frac{1}{t}(s_{11}-x_1)^2- \frac{1}{t}(s_{21}-x_1)^2+
  \frac{1}{t}(is_{12}-x_1)^2\right)\cr & &
  \qquad\qquad\left((is_{12}-s_{11})+(is_{12}-s_{21})\right)
     \widetilde{B}_{21}(s) \, Z^{(0)}_1(s)\, d[s] \ ,
\label{op3}
\end{eqnarray}
where the Berezinian is given by Eq.~(\ref{325}) for $k_1=2$ and
$k_2=1$.  This result is exact for an arbitrary initial condition and
for arbitrary $N$. In the case of a diagonal matrix $H^{(0)}$ as the
initial condition, we have
\begin{eqnarray}
Z^{(0)}_1(s)=\int d[H^{(0)}]
P(H^{(0)}) \prod_{n=1}^N \frac{\prod_{j=1}^{k}(is_{j2}-H^{(0)}_{nn})}
{\prod_{j=1}^{2k}(s_{j1}+i\varepsilon-H^{(0)}_{nn})^{1/2}} 
\label{op4}
\end{eqnarray}
and analogously for the GSE.

In the limit $t\to \infty$ the stationary distribution of classical
Gaussian random matrix theory is recovered. This can be seen by
re--writing Eq.~(\ref{op2}) for the rescaled energy
$\widetilde{x}_1=x_1/t$ and the rescaled source variable
$\widetilde{J}_1=J_1/t$, see also \cite{GUH4}. In this limit the
average over the initial condition yields unity and we arrive at an
integral representation of the one--point correlation function of the
pure GOE,
\begin{eqnarray}
R_1(x_1) &=& \frac{1}{(2\pi)^{3/2}}
  \Im \int\exp\left(-(s_{11}-x_1)^2-(s_{21}-x_1)^2+
  (is_{12}-x_1)^2\right) \cr 
& &\qquad \frac{|s_{11}-s_{21}|}
    {(is_{12}-s_{11})(is_{12}-s_{21})}
    \left(\frac{1}{is_{12}-s_{11}}+\frac{1}{is_{12}-s_{21}} \right)\cr
& &\qquad\qquad \frac{(is_{12})^{N}}
  {(s_{11}+i\varepsilon)^{N/2}(s_{21}+i\varepsilon)^{N/2}}\;
  d[s]\quad\cr & &
\label{op6}
\end{eqnarray}
where the symbol $\Im$ denotes the imaginary part. 
Eq.~(\ref{op6}) is equivalent to the classical
expressions for the one--point functions as in Mehta's book~\cite{MEH}.

Finally, we state an integral expression for the one--point function
for the case of a Poissonian initial conditions, see Eq.~(5.1)
of Ref.~\cite{GUH4}. We have
\begin{equation}
Z^{(0)}_1(s)=\left(\int dz p(z)
                  \frac{\prod_{j=1}^{k}(is_{j2}-z)}
                  {\prod_{j=1}^{2k}(s_{j1}+i\varepsilon-z)^{1/2}}\right)^N \ .
\label{op7b}
\end{equation}
Inserting this initial condition into Eq.~(\ref{op3}) yields the level
density of a transition ensemble between Poisson regularity and GOE in
terms of a fourfold integral. A further analysis will be published
elsewhere.

\subsection{Two--point function}
\label{twop}

The result~(\ref{p413}), derived in Sec.~\ref{subs323}, for the
supermatrix Bessel function $\Phi_{44}(-is,r)$ gives, with the
replacement $r\longrightarrow (x+J)$ and $s\longrightarrow s/t$ and
according to Eq.~(\ref{asym4}), the diffusion kernel for the
two--point function
\begin{equation}
\Gamma_k(s,x+J,t) \ = \ 
 \exp\left(-\frac{1}{t}\left(\trg s^2+\trg(x+J)^2\right)\right)
                   \, \Phi_{44}(-2is/t,x+J) \ .
\label{topf2}
\end{equation}
The derivative with respect to the source terms leaves us with the
two--point correlation function
\begin{eqnarray}
\lefteqn{\widehat{R}_2(x_1,x_2,t)=\frac{2^8\widehat{G}_{44}}
                                       {t^4\pi^2}
                    \int\Biggl( \widetilde{B}_{42}(s) 
               Z_2^{(0)}(s)\frac{\prod_{k=1}^4(is_{12}-s_{k1})(is_{22}-s_{k1})}
                          {(is_{12}-is_{22})^2}\Biggr.}\cr
           &&       \exp\left(-\frac{1}{t}\left(\tr s_1^2+2x_1^2+2x_2^2
                        -2(is_{12}-x_1)^2-2(is_{22}-x_2)^2\right)\right)\cr
           &&       \sum_{k,j}\left[
                    \frac{1}{(is_{12}-s_{k1})(is_{22}-s_{j1})}
                    \left(x_1-t\;\frac{\partial}{\partial s_{j1}}\right)
                    \left(x_2-t\;\frac{\partial}{\partial s_{k1}}\right)
                    +\right.\cr 
          &  &      \frac{t}{(x_1-x_2)(is_{12}-is_{22})
                    (is_{12}-s_{k1})(is_{22}-s_{j1})}
                    \left(x_1-t\;\frac{\partial}{\partial s_{j1}}\right)
                    \left(x_2-t\;\frac{\partial}{\partial s_{k1}}\right)-\cr
         &  &       \left.\frac{1}{2}\frac{t^3}
                  {(x_1-x_2)(is_{12}-s_{k1})
                  (is_{22}-s_{j1})(is_{12}-is_{22})^2}
                    \left(\frac{\partial}{\partial
                        s_{j1}}-\frac{\partial}{\partial
                       s_{k1}}\right)\right]\cr
         &  &  \qquad\Biggl.\Phi_4^{(1)}(-2is_1/t,x)\Biggr)\;d[s] 
                    \, + \, \left( x_1 \longleftrightarrow x_2\right) \  ,
\label{topf3}
\end{eqnarray}
The last line indicates that the integral with $x_1$ and $x_2$
interchanged has to be added.  This yields yet another simplification,
since all terms in Eq.~(\ref{topf3}) antisymmetric under interchange
of $x_1$ and $x_2$ drop out. We arrive at the expression
\begin{eqnarray} 
\lefteqn{R_2(x_1,x_2,t) = \frac{2^8\widehat{G}_{44}}{t^4\pi^2} \; 
   \Im \int \sqrt{\widetilde{B}_{42}(s)}\sqrt{|\Delta_4(s_1)|}
    Z_2^{(0)}(s)}\cr 
     & & \qquad \exp\left(-\frac{1}{t}\left(\tr s_1^2
                            +2x_1^2+2x_2^2
  -2(is_{12}-x_1)^2-2(is_{22}-x_2)^2\right)\right)\cr 
     & &
   \sum_{k,j}\left[ \frac{1}{(is_{12}-s_{k1})(is_{22}-s_{j1})}
    \left(x_1-t\frac{\partial}{\partial s_{j1}}\right)
      \left(x_2-t\frac{\partial}{\partial s_{k1}}\right)
      \right]\cr
     & & \qquad\qquad \Phi_4^{(1)}(-2is_1/t,x) d[s]
\label{topf4}
\end{eqnarray}
The symbol $\Im$ denotes a certain linear combination of
$\widehat{R}_2(x_1,x_2,t)$ as explained in Ref.~\cite{GUH4}.  The
normalization constant obtains from Eq.~(\ref{asym6}) and is given by
$\widehat{G}_{44}=2(2\pi)^{-4}$. This result is an exact expression
for the two--point function of Dyson's Brownian motion for every
initial condition. Plugging in the initial condition of Eq.~(\ref{op7b}),
we find an integral representation of the two--point function for the
transition towards the GOE. We notice that $\Phi_4^{(1)}(-2is_1/t,x)$
is, according to Eq.~(\ref{H9}), given as a double integral. 

As already discussed in Ref.~\cite{GUH4}, the kernels for the
supersymmetric version of Dyson's Brownian motion are the same on all
energy scales. Thus, the integral representation derived here for the
two--level correlation function is, apart from the initial condition,
the same on the so--called unfolded scale which is relevant for
physics applications. The initial condition on the unfolded scale is
found along the lines given in Ref.~\cite{GUH4}.

\section{Summary and Conclusion}
\label{sum4}

We extended the recursion formula of Ref.~\cite{GK1} to superspace.
Due to the group structures in superspace, we could restrict ourselves
to the unitary orthosymplectic supergroup.  As in the ordinary case,
the recursion formula is an exact map of a group integration onto an
iteration in the radial space.  We used it to calculate explicit
expression for certain supermatrix Bessel functions.
  
In ordinary space, we saw that the matrix Bessel functions are only
special cases of the radial functions~\cite{GK1}. We have not yet
studied this further, but in our opinion a similar generalization is
likely to also exist in superspace.

It is a major advantage of the radial Gelfand--Tzetlin coordinates in
superspace that the Grassmann variables appear only as moduli squared.
Thus, the number of Grassmann integrals is a priori reduced by half.
As we showed in detail, this is highly welcome feature for explicit
calculations.  As a remarkable consequence of this recursive way to
proceed, the structure of the supermatrix Bessel functions is only
very little influenced by the matrix dimension. We also saw that the
basic structures of the supermatrix Bessel function for smaller matrix
dimensions survive during the iteration to higher ones.  The matrix
Bessel functions in ordinary space show similar features.  There, the
structure of the matrix Bessel functions is much more influenced by
the group parameter $\beta$ than by the matrix dimension. However, as
in ordinary space, it remains a challenge to find the structure of
these functions for arbitrary matrix dimension.
 
In interesting feature occured which sheds light on the general
properties of the recursion.  Total derivatives showed up in the
integral over the commuting variables after having done the Grassmann
integration. Since similar terms already occured in ordinary space,
they are likely to be an intrinsic property of the recursion formula.
Here, we succeeded in constructing a series of operator identities to
remove them. This was a crucial step for the application of the
recursion formula.  A deeper understanding of these identities is
highly desired.

It should be emphasized that the total derivatives are no boundary
terms in the sense of Rothstein. We showed in detail that such terms
cannot occur because we always work in a compact space.  Thus,
according to a theorem due to Berezin~\cite{BER}, the transformation
of the invariant measure to our radial Gelfand--Tzetlin coordinates
cannot yield Rothstein boundary terms. However, if further integration
over the eigenvalues is required, such terms can emerge.

As an application, we worked out some kernels for the supersymmetric
analogue of Dyson's Brownian Motion.

The radial Gelfand--Tzetlin coordinates are the natural coordinate
system for the matrix Bessel functions in superspace. This
parametrization represents the appropriate tool for the recursive
integration of Grassmann variables. Once the particular features of
this parametrization are better understood, they may allow for the
evaluation of higher dimensional group integrals.

\section*{acknowledgement}
  We thank B.~Balantekin and Z.~Pluhar for useful discussions.  We
  acknowledge financial support from the Deutsche
  Forschungsgemeinschaft, TG for a Heisenberg fellowship and HK for a
  doctoral grant, HK also thanks the Max--Planck-Institute for
  financial support.

\appendix
    
\section{ Radial Gelfand--Tzetlin Coordinates for the 
                     Unitary Orthosymplectic Group ${\it UOSp}(\k_1/2\k_2)$}
\label{appCC1}
\renewcommand{\theequation}{A.\arabic{equation}}
\setcounter{equation}{0}
 
We wish to express the moduli squared of the elements of an orthogonal
$(k_1/2k_2)$ dimensional unit supervector in radial Gelfand--Tzetlin
coordinates.  To illustrate the mechanism, we start with the smallest
non--trivial case, the group ${UOSp}(2/4)$. We notice that there are
at first sight minor, yet crucial, differences to the calculation in
Ref.~\cite{GGT} where we also started with the smallest non--trivial
case. The set of solutions of the Gelfand-Tzetlin equations
(\ref{ss10}) involves one bosonic and two fermionic eigenvalues. The
eigenvalue equations read
\begin{eqnarray}
1 & = & \sum_{p=1}^{2} \left(|v_p^{(1)}|^2 \, 
         +|\alpha_p^{(1)}|^2 \right) ,\label{CC1}\\
0 & = & \sum_{q=1}^{2}\left(
        \frac{|v_q^{(1)}|^2}{s_{q1}-s_{1}^{(1)}} +  
        \frac{|\alpha_q^{(1)}|^2}
             {is_{q2}-s_{1}^{(1)}}\right)\quad ,\label{CC2}\\
z_1 & = & is_{2}^{(1)}
          \prod_{q=1}^{2}\frac{
          (s_{q1}-is_{2}^{(1)})}
          {(is_{q2}-is_{2}^{(1)})^2}
          \sum_{q=1}^2\left(  
          \frac{|v_q^{(1)}|^2}{s_{q1}-is_{2}^{(1)}} +  
         \frac{|\alpha_q^{(1)}|^2}{is_{q2}-is_{2}^{(1)}}\right) \ ,
\label{CC3}
\end{eqnarray}
where the last equation has to be solved in the limit $z_1\to\infty$.
The bosonic equation (\ref{CC2}) has a unique solution 
$s_{1}^{(1)}=s_{11}^\prime$. Taking $s_{11}^\prime$ as new
parameter, Eqs.~(\ref{CC1}) and~(\ref{CC2}) can be solved,
\begin{equation}
|v_p^{(1)}|^2=\frac{s_{p1}-s_{11}^\prime}{s_{p1}-s_{q1}}
            \left(1-\sum_{k=1}^2\frac{is_{k2}-s_{q1}}{is_{k2}-s_{11}^\prime}
            |\alpha_k^{(1)}|^2\right)\quad p = 1,2 \quad .
\label{CC4}
\end{equation} 
We insert these relations in Eq.~(\ref{CC3}) and obtain
\begin{equation}
z_1 \ = \ is_{2}^{(1)}(s_{11}^\prime-is_{2}^{(1)})
          \prod_{q=1}^{2}\frac{
          (s_{q1}-is_{2}^{(1)})}
          {(is_{q2}-is_{2}^{(1)})^2}
          \left(1+\sum_{k=1}^2 \frac{c_k}{is_{k2}-is_{2}^{(1)}}
          |\alpha_k^{(1)}|^2\right) \, 
\label{CC5}
\end{equation}
with $z_1\to\infty$. Here, we have defined the commuting variables
\begin{equation}
c_k \ = \ \frac{\prod_{q=1}^2(is_{k2}-s_{q1})}
           {is_{k2}-s_{11}^\prime} \ , \quad k=1,2 \ .
\label{CC6}
\end{equation}
It remains to determine the set of solutions of the fermionic
eigenvalue equation (\ref{CC5}). To this end, both sides are inverted
\begin{equation}
0 = \prod_{q=1}^2(is_{q2}-is_{2}^{(1)})^2
   \left(1-\sum_{k=1}^2 \frac{c_k}{is_{k2}-is_{2}^{(1)}} 
         |\alpha_k^{(1)}|^2+
         2\prod_{k=1}^2 \frac{c_k}{is_{k2}-is_{2}^{(1)}} 
         |\alpha_k^{(1)}|^2\right) \ .
\label{CC7}
\end{equation}
We can now take the square root on both sides
\begin{equation}
0 = \prod_{q=1}^2(is_{q2}-is_{2}^{(1)})
  \left(1-\frac{1}{2}\sum_{k=1}^2 
  \frac{c_k}{is_{k2}-is_{2}^{(1)}}|\alpha_k^{(1)}|^2+
  \frac{3}{4}\prod_{k=1}^2\frac{c_k}{is_{k2}-is_{2}^{(1)}} 
  |\alpha_k^{(1)}|^2\right) \ .
\label{CC8}
\end{equation}
The most general form of the fermionic eigenvalue is
\begin{equation}
is_{2}^{(1)}= a_{0}+\sum_{k=1}^2 a_{k}|\alpha_k^{(1)}|^2
         +a_{12}\prod_{k=1}^2|\alpha_k^{(1)}|^2 \quad . 
\label{CC9}
\end{equation}
After inserting this ansatz in Eq.~(\ref{CC8}), we obtain two sets of
solutions for the coefficients $a_{i0},a_{i12}$ and $a_{ij}$ with
$i=1,2, j=1,2$
\begin{eqnarray}
is_{12}^\prime&=& is_{12}+\left(c_1+
                 \frac{c_1 c_2}{is_{12}-is_{22}}|\alpha_2^{(1)}|^2\right)
                 \frac{|\alpha_1^{(1)}|^2}{2}\cr 
is_{22}^\prime&=& is_{22}+\left(c_2+
                 \frac{c_1 c_2}{is_{22}-is_{12}}|\alpha_1^{(1)}|^2\right)
                 \frac{|\alpha_2^{(1)}|^2}{2}\qquad .
\label{CC10}
\end{eqnarray}
Remarkably, we have $a_{12}=a_{21}=0$. This allows us to write the
nilpotent part of $is_{k2}^{\prime}$ as the modulus squared of a
new anticommuting coordinate.
\begin{equation}
is_{k2}^\prime= is_{k2} + |\xi_k^\prime|^2\quad .
\label{CC10a}
\end{equation}
We solve the equations (\ref{CC10}) for $|\alpha_p^{(1)}|^2$, insert
the results in Eq.~(\ref{CC4}) and arrive at 
\begin{eqnarray}
|v_p^{(n)}|^2 &  =  &
            \frac{(s_{p1}-s_{11}^{\prime}) 
            \prod_{n=1}^{2}(s_{p1}-is_{n2})^2}
           {(s_{p1}-s_{q1})
           \prod_{n=1}^{2}(s_{p1}-is_{n2}^{\prime})^2} \cr
|\alpha_p^{(n)}|^2 & = & 2\,(is_{p2}^{\prime}-is_{p2})
                           \frac{(is_{p2}-s_{11}^{\prime})
                           (is_{p2}-is_{q2})^2}
                           {(is_{p2}-is_{q2}^{\prime})^2
                           \prod_{n=1}^2 (is_{p2}-s_{n1})}\quad ,\cr
                   &    & \qquad \qquad \qquad\qquad \qquad  
                          \qquad p,q = 1,2 \ , \quad q\neq p\ . 
\label{CC11}
\end{eqnarray}
The structure of Eq.~(\ref{CC11}) indicates the form of the solutions
for groups of higher order as they were stated in Eq.~(\ref{ss5}). They are
checked by inserting them directly into the Gelfand Tzetlin 
equations (\ref{ss10}). The algebra needed is, although tedious, 
straightforward and similar to the one here.

\section{Derivation of Formula~(\ref{LEM1})}
\label{derilem1}
\renewcommand{\theequation}{B.\arabic{equation}}
\setcounter{equation}{0}

The technique we use is an extension of the one developed in App.~D of
Ref.~\cite{GK1}.  First, we rewrite the integral in terms of
$\Theta$ functions. The left hand side reads
\begin{equation}
L_m(s) \int\mu_B(s^\prime,s)\;f(s_1^\prime)\;d[s_1^\prime] 
           \prod_{k\le l}\Theta(s_{k1}-s_{l1}^\prime)
           \prod_{l< n}\Theta(s_{l1}^\prime-s_{n1})\quad .
\label{lde7a}
\end{equation}
The integration domain is now the real axis for all 
variables. The action of $L_m(s)$ on the integral  yields:
\begin{eqnarray}
& &\qquad\int \left(\mu_B(s^\prime,s)\;\sum_{i=1}^{k_1}\sum_{j=1}^{k_1-1}
           \frac{1}{2}\frac{-1}{(is_{12}-s_{i1})(s_{i1}-s_{j1}^\prime)} 
           f(s_1^\prime)\right.\cr
& &\qquad\left.\qquad\qquad\prod_{k\le l}\Theta(s_{k1}-s_{l1}^\prime)
           \prod_{l< n}\Theta(s_{l1}^\prime-s_{n1})\right)\;d[s_1^\prime] \cr
& &\qquad+\int \mu_B(s^\prime,s)\;\sum_{i=1}^{k_1}\frac{1}{is_{m2}-s_{i1}}
            \frac{\partial}{\partial s_{i1}}
            \prod_{k\le l}\Theta(s_{k1}-s_{l1}^\prime)
            \prod_{l< n}\Theta(s_{l1}^\prime-s_{n1})\;d[s_1^\prime]\
            .\cr
         &&
\label{lde8}
\end{eqnarray}
We decompose the first term in partial fractions and find
\begin{eqnarray}
& &\qquad\int\left(\mu_B(s^\prime,s)\;\sum_{i=1}^{k_1}\sum_{j=1}^{k_1-1}
           \frac{1}{2}\frac{-1}{(is_{12}-s_{i1})(is_{12}-s_{j1}^\prime)}\; 
           f(s_1^\prime)-\right.\cr
& &\qquad\left.\Delta_{k_1}(s_1^\prime)\;f(s_1^\prime)\sum_{j=1}^{k_1-1}
           \frac{1}{is_{12}-s_{j1}^\prime}\frac{\partial}{\partial s_{j1}^\prime}
           \frac{1}{\sqrt{-\prod_{i=1}^{k_1}(s_{i1}-s_{j1}^\prime)}}\right)\cr
& &\qquad\qquad\qquad\qquad\prod_{k\le l}\Theta(s_{k1}-s_{l1}^\prime)
           \prod_{l< n}\Theta(s_{l1}^\prime-s_{n1})\;d[s_1^\prime]\cr
& &\qquad+\int \mu_B(s^\prime,s)\;f(s^\prime)\cr
& &\qquad \sum_{i=1}^{k_1}\frac{1}{is_{m2}-s_{i1}}
            \frac{\partial}{\partial s_{i1}}
            \prod_{k\le l}\Theta(s_{k1}-s_{l1}^\prime)
            \prod_{l< n}\Theta(s_{l1}^\prime-s_{n1})\;d[s_1^\prime] \ .
\label{lde9}
\end{eqnarray}
An integration by parts yields
\begin{eqnarray}
& &\qquad\int\;\mu_B(s^\prime,s)\;\sum_{j=1}^{k_1-1}
           \left(-M_{mj}(s_1^\prime,s_1)\right)
           \prod_{k\le l}\Theta(s_{k1}-s_{l1}^\prime)
           \prod_{l< n}\Theta(s_{l1}^\prime-s_{n1})\;d[s_1^\prime]+\cr
& &\qquad\int \mu_B(s^\prime,s)\;f(s^\prime)\left(
            \sum_{i=1}^{k_1}\frac{1}{is_{m2}-s_{i1}}
            \frac{\partial}{\partial s_{i1}}+
            \sum_{j=1}^{k_1-1}\frac{1}{is_{m2}-s_{j1}^\prime}
            \frac{\partial}{\partial s_{j1}^\prime}\right)\cr
& &\qquad\prod_{k\le l}\Theta(s_{k1}-s_{l1}^\prime)
            \prod_{l< n}\Theta(s_{l1}^\prime-s_{n1})\;d[s_1^\prime]\ .
\label{lde9a}
\end{eqnarray}
The derivatives of the $\Theta$ functions yield $\delta$ distributions.
Upon in\-te\-gra\-tion of the $\delta$ distribution the two terms in
the last integral cancel each other. Hence the last term vanishes
identically. This completes the proof.

\section{Alternative Derivations of $\Phi_{22}(\s,\r)$}
\label{altder}
\renewcommand{\theequation}{C.\arabic{equation}}
\setcounter{equation}{0}

We present two different alternative derivations. 
We do this in some detail, because the calculations give 
helpful informations on the r\^ole played by the radial
Gelfand--Tzetlin coordinates.

First, we use a angular parametrization of the 
coset ${\it UOSp}(2/2) /{\it UOSp}(1/2)$ by writing the first column
of $u\in{\it UOSp}(2/2)$ as
\begin{equation}
u_1=\left(\matrix{\sqrt{1-|\alpha|^2}\cos\vartheta\cr
                  \sqrt{1-|\alpha|^2}\sin\vartheta\cr
                  \frac{1}{\sqrt{2}}\alpha\cr
                  \frac{1}{\sqrt{2}}\alpha^*}\right)\quad .
\label{lde12}
\end{equation}
This is a canonical way to parametrize the supersphere $S^{1|2}$ that
is isomorphic to the coset ${\it UOSp}(2/2)/{\it UOSp}(1/2)$, see
Ref.~\cite{ZIR1}.  It coincides with a special choice of the {\it
  angular} Gelfand-Tzetlin coordinates. The invariant measure is in
these coordinates simply $d\mu(u_1)=d\alpha^*d\alpha d\vartheta$.
Thus, one directly obtains the the volume $V(S^{1|2})=0$, see
Ref.~\cite{ZIR1}.  In the parametrization of the measure by {\it
  radial} Gelfand--Tzetlin coordinates (\ref{lde3}), one has to
perform the Grassmann integration and to apply formula~(\ref{LEM1}) to
achieve this result.

Although we use a different coordinate system, we still take advantage
of the recursion formula~(\ref{susyrec}).  To use it in the
parametrization (\ref{lde12}), one has to solve the Gelfand--Tzetlin
equations~(\ref{ss8}) to~(\ref{ss10}) for the eigenvalues. The unique
solution of the bosonic equation (\ref{ss9}) is
\begin{equation}
s_{11}^\prime=a_0+\frac{\prod_{i=1}^2(s_{i1}-a_0)}{is_{12}-a_0}
             |\alpha|^2\quad ,
\quad a_0=\frac{s_{11}+s_{21}}{2}-\frac{s_{11}-s_{12}}{2}
          \cos2\vartheta\quad . 
\label{lde13}
\end{equation}
The fermionic equation yields
\begin{equation}
is_{12}^\prime=is_{12}+\frac{\prod_{i=1}^2(s_{i1}-is_{12})}{is_{12}-a_0}
              |\alpha|^2  \quad .
\label{lde14}
\end{equation}
After inserting Eqs.~(\ref{lde13}) and~(\ref{lde14}) and the measure
$d\mu(u_1)$ into the recursion formula (\ref{susyrec}), the Grassmann
integration can be performed.  Remarkably, we arrive at the
denominator--free expression
\begin{eqnarray}
\Phi_{22}(s,r)&=&\widehat{G}_{22}
                 \int_0^{2\pi}d\vartheta
                 \exp\left(\trg rs -\frac{z}{2}+ 
                 \frac{z}{2}\cos2\vartheta\right)\cr
              & &\Bigg[
                 \left(\prod_{i=1}^2(r_{i1}-ir_{12})(s_{i1}-is_{12})
                 +\frac{1}{2}
                 \sum_{i=1 \atop j=1}^2(s_{j1}-is_{12})(r_{i1}-ir_{12})
                 \right)\Bigg.\cr
              & &\qquad-\frac{1}{2}\left(ir_{12}-\frac{1}{2}(r_{11}+r_{21})\right)
                 \left(s_{11}-s_{21}\right)\cos2\vartheta -\cr
              & &\qquad\Bigg.\frac{z}{8}(ir_{12}-r_{21})(s_{11}-s_{21})
                 \sin^22\vartheta\Bigg]\ .   
\label{lde15}
\end{eqnarray}
To make contact with Eq.~(\ref{lde11}) one has to realize, that 
in Eq.~(\ref{lde15})
an additional total derivative appears in the integrand.
This becomes obvious if one adds and subtracts $z/4\cos2\vartheta$
in the squared bracket of Eq.~(\ref{lde15})
\begin{eqnarray}
\Phi_{22}(s,r)&=&\widehat{G}_{22}
                 \int_0^{2\pi}d\vartheta
                 \exp\left(\trg rs - \frac{z}{2}+ 
                  \frac{z}{2}\cos2\vartheta\right)\cr
              & &\biggl(\prod_{i=1}^2(r_{i1}-ir_{12})(s_{i1}-is_{12})
                                \biggr.\cr
              & &\qquad\biggl.
                 +\frac{1}{2}\sum_{i=1\atop j=1}^2(s_{j1}-is_{12})(r_{i1}-ir_{12})
                 -\frac{z}{4}\cos2\vartheta\biggr)\cr
              & &+\frac{ir_{12}-r_{21}}{r_{21}-r_{11}}
                 \int_0^{2\pi}d\vartheta
                 \frac{\partial^2}{\partial\left(2\vartheta\right)^2}  
                 \exp\left(\trg rs -
                 \frac{z}{2}\left(1-\cos2\vartheta\right)\right)\ .
\label{lde16}
\end{eqnarray}
While the first integral reproduces Eq.~(\ref{lde11}), the second one
vanishes identically. In general, in performing Grassmann
integrations, one has to take care of boundary contributions
\cite{BER,ROT}.  These contributions can appear whenever even
coordinates are shifted by nilpotents and the function one integrates
does not have compact support \cite{BER}. However, in our case the
base space is always given by a $n$ dimensional sphere, i.~e.~by a
compact manifold {\it without boundary}.  Thus in a properly chosen
coordinate system, no boundary terms should appear. With regard to
Eq.~(\ref{lde16}) this means: the fact that the last term vanishes, is
a direct consequence of the compactness of the circle and of the
analyticity of the function, that we integrate.  However, in the
radial Gelfand--Tzetlin coordinates, only the moduli squared of the
vector $u_1$ are determined. Therefore, not the whole sphere, but only
a $(2^{n+1})^{\rm th}$ segment of it is covered by Eq.~(\ref{ss5}).
In our case, not the circle but only a quarter of it is parametrized.
This is allowed since the supermatrix Bessel functions depend only on
the moduli squared $|u_{i1}|^2$.  Nevertheless, one has to ensure that
the introduction of these artificial boundaries does not alter the
result.  To this end we use the following integration formula.

Let $s_{11}<s_{11}^\prime<s_{21}$ be real and let
$\xi^\prime,\xi^{\prime*}$ be anticommuting. Furthermore, define
\begin{equation}
f(s_{11}^\prime,\xi,\xi^{*})=
f_0(s_{11}^\prime)+f_1(s_{11}^\prime)|\xi|^2\quad , 
\label{lde17}
\end{equation}
with two analytic functions $f_0(s_{11}^\prime),f_1(s_{11}^\prime)$. 
Then the integral
\begin{equation}
I=\int_{s_{11}}^{s_{21}} ds_{11}^\prime d\xi^* d\xi f(s_{11}^\prime,\xi,\xi^*)
\label{lde18}
\end{equation}
transforms under a shift of $s_{11}^\prime$ by nilpotents 
\begin{equation}
s_{11}^\prime=y+g(y)|\xi|^2
\label{lde19}
\end{equation}
in the following way,
\begin{equation}
I=\int_{s_{11}}^{s_{21}} dy d\xi^* d\xi \frac{\partial s_{11}^\prime}{\partial y}
  f\left(y(s_{11}^\prime),\xi,\xi^*\right) - 
  \left[f_0(s_{21})g(s_{21})-f_0(s_{11})g(s_{11})\right] \ .
\label{lem2}
\end{equation}
The proof is by direct calculation. The second term in
Eq.~(\ref{lem2}) is often referred to as boundary term. It can be
viewed as the integral of a total derivative, i.e.~an exact one--form,
that has to be added to the integration measure for functions with
non--compact support~\cite{ROT}. For functions of an arbitrary number
of commuting and anticommuting arguments, a similar integral formula
holds with additional boundary terms~\cite{BER}.  In going from the
canonical coordinates $(\vartheta,\alpha,\alpha^*)$ to the radial
ones $(s_{11}^\prime,\xi_1^\prime,\xi_1^{\prime*})$, in principle
boundary terms can arise, since the bosonic Gelfand--Tzetlin
eigenvalue (\ref{lde13}) contains nilpotents. However, the crucial
quantity is $g(y)$ in formula~(\ref{lem2}) which, in our case, is
given by
\begin{equation}
g(s_{11}^\prime)=\frac{\prod_{i=1}^2(s_{i1}-s_{11}^\prime)}
                     {is_{12}-s_{11}^\prime} \ .
\label{lde21}
\end{equation}
Thus, $g(s_{11}^\prime)$ causes the boundary term to vanish at
$s_{11}$ and $s_{21}$. It is the product structure of the left hand
side of Eq.~(\ref{ss5}) which always guarantees the vanishing of the
boundary terms, when one goes from the Cartesian set of coordinates to
the radial Gelfand--Tzetlin coordinates. 

Therefore, one may think of the denominators, arising in
Eqs.~(\ref{lde4}),~(\ref{lde5}), as 
belonging to total derivatives of functions, which vanish
at the boundaries. Keeping this in mind we derive 
Eq.~(\ref{lde11}) in yet another way. We expand the product
\begin{equation}
\prod_{q=1}^{k_1}\left(is_{m2}-s_{q1}\right) = 
\sum_{n=0}^{k_1} \frac{1}{n!}\ (is_{m2}-s_{j1}^\prime)^n
             \frac{\partial^n}{\partial (s_{j1}^\prime)^n}
            \prod_{q=1}^{k_1}\left(s_{j1}^\prime-s_{q1}\right)\quad ,
\label{lde23}
\end{equation}
and insert it into the integral
\begin{eqnarray}
\lefteqn{\int_{s_{11}}^{s_{21}}\!\ldots\!\int_{s_{(k_1-1)1}}^{s_{k_11}}\!
\mu_B(s_1^\prime,s_1) K_{mj}(s_1^\prime,s_1)f(s_1^\prime)d[s_1^\prime]=}\cr
&&\int_{s_{11}}^{s_{21}}\!\ldots\!\int_{s_{(k_1-1)1}}^{s_{k_11}}\!
             \mu_B(s_1^\prime,s_1)
  \prod_{n=1}^{k_1}\left(is_{m2}-s_{n1}\right)
  M_{mj}(s_1^\prime,s_1)\,f(s_1^\prime)\,d[s_1^\prime] \ . \cr
&&
\label{lde24}
\end{eqnarray}
We can remove the term proportional to $(is_{m2}-s_{j1}^\prime)^{-2}$
in the integrand by an integration by parts. Through the expansion
(\ref{lde23}), the vanishing of the boundary terms is assured. We
arrive at
\begin{eqnarray}
K_{mj}(s_1^\prime,s_1)&=&-\sum_{n=2}^{k_1} 
                        \frac{1}{n!}(is_{m2}-s_{j1}^\prime)^{n-2}
                        \frac{\partial^n}{\partial (s_{j1}^\prime)^n}
                        \prod_{q=1}^{k_1}\left(s_{j1}^\prime-s_{q1}\right)+\cr
                     & &\frac{\prod_{q=1}^{k_1}\left(is_{m2}-s_{q1}\right)-
                        \prod_{q=1}^{k_1}\left(s_{j1}^\prime-s_{q1}\right)}
                           {is_{m2}-s_{j1}^\prime}\cr
                     & &\left(\frac{1}{2}\sum_{q=1}^{k_1}
                        \frac{1}{is_{m2}-s_{q1}}
                        -\frac{1}{2}\sum_{q=1}^{k_1} 
                        \frac{1}{s_{j1}^\prime-s_{q1}}
                        -\sum_{q=1 \atop q \neq j}^{k_1}
                        \frac{1}{s_{j1}^\prime-s_{q1}^\prime}-
                        \frac{\partial}{\partial s_{j1}^\prime}\right)\ .\cr
                     & &
\label{lde25}
\end{eqnarray}
We notice that in the new operator $K_{mj}(s_1^\prime,s_1)$ all
denominators of the type $(is_{m2}-s_{j1}^\prime)^{-1}$ have
disappeared. For $k_1=2$, we calculate
\begin{equation}
K_{11} = -\left(is_{12}+s_{11}^\prime-s_{11}-s_{21}\right)
          \frac{\partial}{\partial s_{q1}^\prime} \quad ,
\label{lde26}
\end{equation}
which can be inserted into Eq.~(\ref{lde4}) by using the definition
(\ref{lde24}). Finally, the result (\ref{lde11}) is reproduced by the
substitution
\begin{equation}
s_{11}^\prime = \frac{s_{11}+s_{21}}{2}-\frac{s_{11}-s_{12}}{2}
               \cos2\vartheta \quad , 
\label{lde27}
\end{equation}
see Eq.~(\ref{lde13}).  In other words, we have seen that the result
of this procedure is summarized in formula~(\ref{LEM1}).

Finally, some remarks are in order: First, from this discussion, one
might conclude that the radial Gelfand--Tzetlin coordinates are less
adapted to the problem than the canonical
parametrization~(\ref{lde12}), because, in the latter, no denominators
appear. We stress that this is not true. Certainly, the denominators
appear due to the shift of the bosonic variable by nilpotents in
Eq.~(\ref{lde13}). However, the difficulty in deriving
Eq.~(\ref{lde11}) is the identification of the different parts of the
integrand after the Grassmann integration.  Some of them belong to
total derivatives and this problem exists in both parametrizations.
Second, we emphasize, that the appearance of total derivatives in the
integrand is not a peculiarity of supersymmetry.  Already in
Ref.~\cite{GK1} where the matrix Bessel functions in ordinary space
were treated we had to solve a similar problem. The appearance of
these total derivatives is an intrinsic property of the recursion
formula. A geometrical interpretation of this phenomenon is highly
desired.

\section{Details for the Derivation of $\Phi_{34}(-\iit\s,\r)$}
\label{appH3}
\renewcommand{\theequation}{D.\arabic{equation}}
\setcounter{equation}{0}
 
We always consider the case that one matrix has an additional
degeneracy according to Eq.~(\ref{p46}) and (\ref{p410}).  We
introduce the notation
\begin{equation}
S_{ij} = (s_{i1}- is_{j2}) \quad {\rm and} \quad
R_{ij}= (r_{i1}- ir_{j2}) \ .
\label{H1a}
\end{equation}
Due to the degeneracy, $\Phi_{24}(-is,\widetilde{r})$ simplifies
enormously as compared to the general result (\ref{p45}).  We insert
it into the recursion formula~(\ref{susyrec}) and do the trivial
integral over the $O(2)$ subgroup.  After performing the Grassmann
integrals we arrive at an expression similar to Eq.~(\ref{p42}),
\begin{eqnarray}
\Phi_{34}(-is,r)&=& 4\;\widehat{G}_{34} 
                   \exp\left(\tr r_2s_2 +
                   r_{11}(s_{11}+s_{21})\right)\cr
                & & \int d\mu_B(s_1^\prime,s_1)
                   \prod_{i=1}^2 R_{1i}\;\prod_{j=1}^3 S_{ji}\cr
                & & \Biggl[
                   \left(\frac{1}{\Delta_2^2(ir_2)\Delta_2^2(is_2)}+
                   \frac{1}{\Delta_2^3(ir_2)\Delta_2^3(is_2)}\right)
                          \Biggr.\cr
                & & \left(4\prod_{i=1}^2 R_{i1}
                   -2\sum_{k=1}^3R_{21}S_{k1}^{-1}
                   +2\sum_{j=1}^2 M_{1j}^\rightarrow(s_1^\prime,s_1)
                   \right)\cr
              & &   \left(4\prod_{i=1}^2 R_{i2}
                   -2\sum_{k=1}^3R_{22} S_{k2}^{-1}
                   +2\sum_{j=1}^2 M_{2j}(s_1^\prime,s_1)\right)
                   \cr 
              & &  +\left(\frac{1}{\Delta_2^2(ir_2)\Delta_2^3(is_2)}+
                   \frac{1}{\Delta_2^3(ir_2)\Delta_2^4(is_2)}\right)\cr
              & &   \sum_{j=1}^2\left(\frac{4}{is_{12}-s_{j1}^\prime}
                   M_{2j}(s_1^\prime,s_1)-\frac{4}{is_{22}-s_{j1}^\prime}
                   M_{1j}(s_1^\prime,s_1)\right)\cr
              &  & -\left(\frac{1}{\Delta_2^2(ir_2)\Delta_2^3(is_2)}+
                   \frac{1}{\Delta_2^3(ir_2)\Delta_2^3(is_2)}\right)
                   \prod_{k=1 \atop j=1}^2
                   \frac{2}{is_{k2}-s_{j1}^\prime}+\cr
              &  & \frac{2}{\Delta_2^3(ir_2)\Delta_2^4(is_2)}
                   \left( \trg r +r_{11}
                   -\sum_{i=1}^2S_{i2}^{-1}\right)
                   \sum_{j=1}^2 M_{1j}(s_1^\prime,s_1)-\cr
              &  & \frac{2}{\Delta_2^3(ir_2)\Delta_2^4(is_2)}
                   \left(\trg r +r_{11}
                   -\sum_{i=1}^2 S_{i1}^{-1}\right)
                   \sum_{j=1}^2M_{2j}(s_1^\prime,s_1)\cr
              &  & \Biggr.-\frac{4}{\Delta_2^3(ir_2)\Delta_2^3(is_2)}
                   \sum_{k=1}^3\prod_{j=1}^2
                   R_{2j}S_{kj}^{-1}\Biggr]\cr
              &  & \qquad
                   \exp\left((s_{21}^\prime+s_{11}^\prime)
                       (r_{21}-r_{11})\right) \cr
              &  & \qquad\qquad
                   \, + \, \left( ir_{12}\longleftrightarrow
                     ir_{22}\right) \ .            
\label{H2}
\end{eqnarray}
Formulae~(\ref{lem3}) and~(\ref{LEM1}) are needed to remove the
denominators, in a way similar as for $\Phi_{24}(s,r)$. A single sum
$\sum_{j=1}^2 M_{1j}(s_1^\prime,s_1)$ transforms according to
formula~(\ref{LEM1}). Moreover, we observe that parts of
Eq.~(\ref{H2}) together with the product $\sum_{j=1}^2
M_{1j}^\to(s_1^\prime,s_1) \sum_{k=1}^2 M_{2k}(s_1^\prime,s_1)$ yield
exactly the integrand of formula~(\ref{lem3}). Thus, it can be
transformed accordingly.  After rearranging terms, we arrive at the
result~(\ref{p49}).

\section{Details for the Derivation of $\Phi_{44}(-\iit\s,\r)$}
\label{appH4}
\renewcommand{\theequation}{E.\arabic{equation}}
\setcounter{equation}{0}

For the recursion, we need $\Phi_{34}(s,r)$ with degenerate
$\widetilde{r}=\diag(r_{11},r_{21},r_{21})$ according to
~Eq.~(\ref{p46}).  Using the integral representation~(\ref{SS1}) for
$\Phi_3^{(1)}(-is_1^\prime,\widetilde{r}_1)$ we find the helpful
identity
\begin{eqnarray}
\lefteqn{\frac{\partial}{\partial s_{i1}^\prime}
\frac{\partial}{\partial s_{j1}^\prime}\exp\left(-r_{21}\tr s_1^\prime\right)
\Phi_{3}^{(1)}(-is_1^\prime,\widetilde{r}_1)=}\cr
&& 
\frac{1}{2}\frac{1}{s_{i1}^\prime-s_{j1}^\prime}\left(
\frac{\partial}{\partial s_{i1}^\prime}-\frac{\partial}{\partial s_{j1}^\prime}
\right)\exp\left(-r_{21}\tr s_1^\prime\right)
\Phi_{3}^{(1)}(-is_1^\prime,\widetilde{r}_1)\qquad .
\label{H4a}
\end{eqnarray}
We stress that this relation, which is crucial in the derivation, only
holds, because of the degeneracy in the matrix $\widetilde{r}_1$.
Employing Eq.~(\ref{SS1}) and another identity,
\begin{eqnarray}
\lefteqn{\sum_{i=1}^3 \frac{\partial}{\partial s_{i1}^\prime}
\exp\left(-r_{21}\tr
  s_1^\prime\right)\Phi_{3}^{(1)}(-is_1^\prime,\widetilde{r}_1)=}\cr
& &\qquad\qquad\qquad(r_{11}-r_{21})\exp\left(-r_{21}\tr s_1^\prime\right)
    \Phi_{3}^{(1)}(-is_1^\prime,\widetilde{r}_1)\quad .
\label{H5}
\end{eqnarray}
We insert $\Phi_{34}(s^\prime,\widetilde{r})$ into the recursion formula~(\ref{susyrec}).
We then can arrange the terms emerging from the Grassmann integration
in a way similar to the former cases. We obtain
\begin{eqnarray}
\lefteqn{\Phi_{44}(-is,r)=4\;\widehat{G}_{44} \exp\left(\tr r_2s_2 +
                   r_{11}\tr s_1\right)
                   \int d\mu_B(s_1^\prime,s_1)
                   \prod_{i=1}^2 R_{2i}\prod_{j=1}^4 S_{ji}}\cr
              &  & \left[\left(\frac{1}{\Delta_2^2(ir_2)\Delta_2^2(is_2)}+
                   \frac{1}{\Delta_2^3(ir_2)\Delta_2^3(is_2)}\right)\right.\cr
              &  & \left(8 R_{21}R_{11}^2-4 R_{11}R_{21} 
                   \sum_{k=1}^4 S_{k1}^{-1}
                   +4\sum_{j=1}^3\left( R_{21}
                   -\frac{\partial^\to}
                   {\partial s_{j1}^\prime}\right) 
                   M_{1j}^\rightarrow(s_1^\prime,s_1)
                   \right)\cr
              &  & \left(8 R_{22}R_{12}^2
                   - 4 R_{12}R_{22} \sum_{k=1}^4 S_{k2}^{-1}
                   +4\sum_{j=1}^3\left( R_{22}-
                   \frac{\partial^\to}
                   {\partial s_{j1}^\prime}\right)
                   M_{2j}(s_1^\prime,s_1)\right)\cr 
              &  &+\frac{16}{\Delta_2^3(ir_2)\Delta_2^4(is_2)}
                   \sum_{j=1}^3 M_{1j}^\to(s_1^\prime,s_1)
                   \left( \frac{1}{2}R_{11}R_{12}\left( \trg r 
                   -\sum_{i=1}^4 S_{i1}^{-1}\right) +\right. \cr
              &  &\left(r_{21}-r_{11}-\frac{\partial}
                   {\partial s_{j1}^{\prime}}\right)\cr
              &  &\left.\left( R_{22}R_{12}+ R_{11}R_{12}+R_{11}R_{21}+
                   \frac{1}{2}(R_{12}+R_{22})\sum_{i=1}^4
                   S_{i1}^{-1}\right)\right)\cr
              &   &-\frac{16}{\Delta_2^3(ir_2)\Delta_2^4(is_2)}
                   \sum_{j=1}^3 M_{2j}^\to(s_1^\prime,s_1)
                   \left(\frac{1}{2} R_{11}R_{12}\left( \trg r 
                   -\sum_{i=1}^4 S_{i2}^{-1}\right) +\right. \cr
              &   &\left(r_{21}-r_{11}-\frac{\partial}
                   {\partial s_{j1}^{\prime}}\right)\cr
              &  & \left.\left(R_{11}R_{21}+ R_{11}R_{12}+R_{12}R_{22}+
                   \frac{1}{2}(R_{12}+R_{22})\sum_{i=1}^4 
                   S_{i2}^{-1}\right)\right)\cr
              &  & \left.-\frac{16}{\Delta_2^3(ir_2)\Delta_2^3(is_2)}
                   \sum_{k=1}^4\prod_{i,j}^2
                   R_{ij}S_{kj}^{-1}\right]
                   \exp\left(-r_{11}\tr s_1^\prime\right)
                   \Phi_3^{(1)}(-is_1^\prime,\widetilde{r}_1)\cr
              &   &\qquad\qquad\qquad 
                 \ + \ C(s,r)  + \ 
                \left( ir_{12}\longleftrightarrow ir_{22}\right)\qquad.
\label{H6}
\end{eqnarray}
Again, all operators with an arrow are understood to act only onto the
term outside the squared bracket, i.e.~onto $\exp\left(-r_{11}\tr
s_1^\prime\right) \Phi_3^{(1)}(-is_1^\prime,\widetilde{r}_1)$.
In the function $C(s,r)$, we summarized the terms that are expected
to arise due to non--commutativity of some operators acting on the
integral and some operators acting under the integral. The last two
lines in formula~(\ref{lem3}) are examples of such terms
\begin{eqnarray}
\lefteqn{C(s,r)= 4\; \widehat{G}_{44} \exp\left(\tr r_2s_2 +
                   r_{11}\tr s_1\right)
                   \int d\mu_B(s_1^\prime,s_1)
                   \prod_{i=1}^2 R_{2i}\prod_{j=1}^4 S_{ji}}\cr
               &  &\Biggl[\left(\frac{1}{\Delta_2^2(ir_2)\Delta_2^3(is_2)}+
                   \frac{1}{\Delta_2^3(ir_2)\Delta_2^4(is_2)}\right)
                                           \Biggr.\cr
               &  &\sum_{j=1}^3\left(R_{11}R_{12}+
                   (R_{11}+R_{12})(r_{21}-r_{11})-
                   (R_{11}+R_{12})\frac{\partial^\to}
                    {\partial s_{j1}^\prime}\right)\cr
              &  & \qquad\qquad\qquad
                   \left(\frac{16}{is_{12}-s_{j1}^\prime}
                   M_{2j}(s_1^\prime,s_1)-\frac{16}{is_{22}-s_{j1}^\prime}
                   M_{1j}(s_1^\prime,s_1)\right)\cr
              &  & -\left(\frac{1}{\Delta_2^2(ir_2)\Delta_2^2(is_2)}+
                   \frac{1}{\Delta_2^3(ir_2)\Delta_2^3(is_2)}\right)\cr
              &  & \prod_{k=1}^2\prod_{j=1}^3\left(R_{11}R_{12}+
                   (R_{11}+R_{12})(r_{21}-r_{11})-(R_{11}+R_{12})
                   \frac{\partial^\to}{\partial s_{j1}^\prime}\right)
                   \frac{8}{is_{k2}-s_{j1}^\prime}\cr
              &  &  -\left(\frac{1}{\Delta_2^2(ir_2)\Delta_2^2(is_2)}+
                   \frac{1}{\Delta_2^3(ir_2)\Delta_2^3(is_2)}\right)\cr
              &  &\qquad\qquad\qquad
                 \sum_{j,k}^3\left((r_{21}-r_{11})-
                   \frac{\partial^\to}{\partial s_{j1}^\prime}\right)
                  \left((r_{21}-r_{11})-
                   \frac{\partial^\to}{\partial s_{k1}^\prime}\right)
                  M_{1j}^\to M_{2k}\cr
              &  & -\frac{8}{\Delta_2^2(ir_2)\Delta_2^3(is_2)}
                   \sum_{j=1}^3\left((r_{21}-r_{11})- 
                   \frac{\partial^\to}{\partial s_{j1}^\prime}\right)
                   \left(\sum_{k=1}^4 S_{k2}^{-1}M_{j1}^\to-
                         \sum_{k=1}^4 S_{k1}^{-1}M_{j2}\right)\cr
              &  & \Biggl.+\frac{8}{\Delta_2^2(ir_2)\Delta_2^3(is_2)}
                   \left(\sum_{i\neq j}M_{j1} M_{j2}^\to
                   \left(\frac{\partial}{\partial s_{i1}^\prime}-
                   \frac{\partial}{\partial
                   s_{j1}^\prime}\right)\right)
                               \Biggr] \cr
              & & \qquad\qquad \exp\left(-r_{11}\tr s_1^\prime\right)
                   \Phi_3^{(1)}(-is_1^\prime,\widetilde{r}_1) \ .
\label{H6aa}
\end{eqnarray}
In order to evaluate Eqs.~(\ref{H6}) and~(\ref{H6aa}) we need some
more properties of the matrix Bessel function
$\Phi_4^{(1)}(-is_1,r_1)$.  We investigate the action of
$\widetilde{L}_k$ on $\Phi_4^{(1)}(-is_1,r_1)$ using the integral
representation~(\ref{H9}).

After a straightforward calculation involving an integration 
by parts we find
\begin{eqnarray}
\lefteqn{\widetilde{L}_k \exp\left(-r_{11}\tr s_1\right)
            \Phi_4^{(1)}(-is_1,r_1) \ = } \cr
   &  &\qquad\qquad \sum_{i=1}^4 \frac{1}{is_{k2}-s_{i1}}
         \left((r_{21}-r_{11})^2 +
         (r_{21}-r_{11})\frac{\partial}{\partial s_{i1}}\right)\cr
   &  &  \qquad\qquad\qquad\qquad
         \exp\left(-r_{11}\tr s_1\right)\Phi_4^{(1)}(-is_1,r_1)
\label{H10}
\end{eqnarray}     
Now Eqs.~(\ref{H6}) and~(\ref{H6aa}) can be enormously simplified by the
observation that
\begin{equation}
\left((r_{21}-r_{11})L_k - \widetilde{L}_k\right)
\exp\left(-r_{11}\tr s_1\right)\Phi_4^{(1)}(-is_1,r_1) = 0\qquad ,
\label{H11}
\end{equation}
which follows directly from Eq.~(\ref{H10}). We find for Eq.~(\ref{H6})
\begin{eqnarray}
\lefteqn{\Phi_{44}(-is,r)= 4\;\widehat{G}_{44} \exp\left(\tr r_2s_2 +
                   r_{11}\tr s_1\right)
                   \int d\mu_B(s_1^\prime,s_1)
                   \prod_{i=1}^2 R_{1i}\prod_{k=1}^4 S_{ki}}\cr
              &  & \Biggl[\left(\frac{1}{\Delta_2^2(ir_2)\Delta_2^2(is_2)}+
                   \frac{1}{\Delta_2^3(ir_2)\Delta_2^3(is_2)}\right)\Biggr.\cr
              &  & \biggl(R_{12}R_{11}
                   \left(8 R_{12}R_{22}
                     - 4 R_{22}\sum_{k=1}^4 S_{k2}^{-1}\right)\biggr.\cr
              &  & \qquad \left(8 R_{11}R_{21}-4 R_{21} 
                   \sum_{k=1}^4 S_{k1}^{-1}
                   +4\sum_{j=1}^3 M_{1j}^\rightarrow(s_1^\prime,s_1)
                   \right) \cr
              &  & +R_{12}R_{11}\left(8 R_{11}R_{21}-4 R_{21} 
                   \sum_{k=1}^4 S_{k1}^{-1}\right) \cr
              &  & \qquad \left(8 R_{22}R_{12}
                   - 4 R_{22}\sum_{k=1}^4 S_{k2}^{-1}
                   +4\sum_{j=1}^3 M_{2j}(s_1^\prime,s_1)\right)\cr
              &  & \qquad\qquad\qquad+\sum_{j,i} R_{11}
                  \left(r_{21}-r_{11}-\frac{\partial^\to}{\partial 
                  s_{j1}^\prime}\right)
                  M_{1i}^\to(s_1^\prime,s_1)M_{2j}(s_1^\prime,s_1)\cr
              &  &\qquad\qquad\qquad+\sum_{j,i} R_{12}
                  \left(r_{21}-r_{11}-\frac{\partial^\to}{\partial 
                  s_{i1}^\prime}\right)
                  M_{1i}^\to(s_1^\prime,s_1)M_{2j}(s_1^\prime,s_1)\cr
              &  &\biggl.\qquad\qquad\qquad+\sum_{j,i} R_{21}R_{22}
                  M_{1j}^\to(s_1^\prime,s_1)M_{2i}(s_1^\prime,s_1)\biggr)\cr
              &   &+\frac{8}{\Delta_2^3(ir_2)\Delta_2^4(is_2)}
                   R_{11}R_{12}\left( \trg r 
                   -\sum_{i=1}^4 S_{i1}^{-1}\right)
                   \sum_{j=1}^3 
                   \left(M_{1j}(s_1^\prime,s_1)-
                   M_{2j}(s_1^\prime,s_1)\right) \cr
              &  & \Biggl.-\frac{16}{\Delta_2^3(ir_2)\Delta_2^3(is_2)}
                   \sum_{k=1}^4\prod_{i,j}^2
                   R_{ij}S_{kj}^{-1}\Biggr]
                   \exp\left(-r_{11}\tr s_1^\prime\right)
                   \Phi_3^{(1)}(-is_1^\prime,\widetilde{r}_1)\cr
              &   &\qquad\qquad\qquad\qquad 
                     \, + \, C(s,r) \, + \, 
                \left(ir_{12}\longleftrightarrow ir_{22}\right) \ .
\label{H12a}
\end{eqnarray}
The terms contained in Eq.~(\ref{H6aa}) simplify, too. We arrive at
\begin{eqnarray}
\lefteqn{C(s,r)= 4\;\widehat{G}_{44} \exp\left(\tr r_2s_2 +
                   r_{11}\tr s_1\right)
                   \int d\mu_B(s_1^\prime,s_1)
                   \prod_{i=1}^2 R_{1i}\prod_{j=1}^4 S_{ji}}\cr
               &  &\left[\left(\frac{1}{\Delta_2^2(ir_2)\Delta_2^3(is_2)}+
                   \frac{1}{\Delta_2^3(ir_2)\Delta_2^4(is_2)}\right)\right.\cr
               &  &\sum_{j=1}^3\left(R_{11}R_{12}+
                   (R_{11}+R_{12})(r_{21}-r_{11})-
                   (R_{11}+R_{12})\frac{\partial^\to}
                    {\partial s_{j1}^\prime}\right)\cr
              &  & \qquad\qquad\qquad
                   \left(\frac{16}{is_{12}-s_{j1}^\prime}
                   M_{2j}(s_1^\prime,s_1)-\frac{16}{is_{22}-s_{j1}^\prime}
                   M_{1j}(s_1^\prime,s_1)\right)\cr
              &  & -\left(\frac{1}{\Delta_2^2(ir_2)\Delta_2^2(is_2)}+
                   \frac{1}{\Delta_2^3(ir_2)\Delta_2^3(is_2)}\right)\cr
              &  & \qquad
                   \prod_{k=1 \atop j=1}^2\left(R_{11}R_{12}+
                   (R_{11}+R_{12})(r_{21}-r_{11})-(R_{11}+R_{12})
                   \frac{\partial^\to}{\partial s_{j1}^\prime}\right)
                   \frac{8}{is_{k2}-s_{j1}^\prime}\cr
              &  & \left.+\frac{8}{\Delta_2^2(ir_2)\Delta_2^3(is_2)}
                   \left(\sum_{i\neq j}M_{j1}^\to(s_1^\prime,s_1)
                    M_{j2}(s_1^\prime,s_1)
                   \left(\frac{\partial}{\partial s_{i1}^\prime}-
                   \frac{\partial}{\partial s_{j1}^\prime}\right)
                   \right)\right]\cr
              &  & \qquad\qquad\qquad\exp\left(-r_{11}\tr s_1^\prime\right)
                   \Phi_3^{(1)}(-is_1^\prime,\widetilde{r}_1)\ . 
\label{H13}
\end{eqnarray}
To further evaluate the expressions, we can now invoke a symmetry
argument between the eigenvalues $r_{11}$ and $r_{21}$, respectively.
Since the product $R_{11}R_{12}$ appears as a prefactor in front of
the integral~(\ref{H12a}), $R_{21}R_{22}$ must also appear as a
prefactor in the final result.  Thus, all terms in Eqs.~(\ref{H12a})
and~(\ref{H13}) which do not contain $R_{21}R_{22}$ as a factor must
yield zero. The remaining terms which are proportional to
$R_{21}R_{22}$ can again be treated using formulae~(\ref{LEM1})
and~(\ref{lem3}). However, we want to show explicitly that this line
of arguing is correct and that the other terms indeed vanish.  To this
end, we need an additional identity to treat the operator product
\begin{equation}
\sum_{j=1}^2 \frac{\partial^\to}{\partial s_{j1}^\prime}
             M_{1j}^\to(s_1^\prime,s_1)
\sum_{k=1}^2 M_{2k}(s_1^\prime,s_1)\qquad .
\label{H6a}
\end{equation}
The required identity is given by the following formula: The same
conditions as for formula~(\ref{LEM1}) apply, furthermore we define
\begin{eqnarray}
L_m^\to(s)\widetilde{L}_n(s)& = &
                 \sum_{i,j}\frac{1}{(is_{m2}-s_{i1})(is_{n2}-s_{j1})}
                 \frac{\partial^3}{\partial s_{i1}\partial s_{j1}^2}+\cr
       &&        \frac{1}{2}\sum_{i,j}
                 \frac{1}{(is_{m2}-s_{i1})(is_{n2}-s_{j1})}\cr
       &&\qquad\qquad \frac{\partial}{\partial s_{i1}}
                 \sum_{k\neq j}^{k_1}\frac{1}{s_{j1}-s_{k1}}
                 \left(
                 \frac{\partial}{\partial s_{j1}}-
                 \frac{\partial}{\partial s_{k1}}\right) \ .\cr
       &&
\label{H14}
\end{eqnarray} 
Then we have
\begin{eqnarray}
\lefteqn{L_m^\to(s)\widetilde{L}_n(s)\int_{s_{11}}^{s_{21}}\!\ldots\! 
        \int_{s_{(k_1-1)1}}^{s_{k_11}}
        \mu_B(s^\prime,s) d[s_1^\prime] f(s_1^\prime) =}\cr 
 &&     \int_{s_{11}}^{s_{21}}\ldots \int_{s_{(k_1-1)1}}^{s_{k_11}}
        \left[\sum_{j=1}^{k_1-1}\sum_{i=1}^{k_1-1}
        M_{mi}^\to(s_1^\prime,s_1)
        \frac{\partial^\to}{\partial s_{j1}^\prime} 
        M_{nj}(s_1^\prime,s_1)f(s_1^\prime)-\right.\cr
 &&     \frac{1}{is_{n2}-is_{m2}}\sum_{i=1}^{k_1-1}\left(
        \frac{1}{is_{m2}-s_{i1}^\prime}
        \frac{\partial^\to}{\partial s_{i1}^\prime}
        M_{ni}(s_1^\prime,s_1)-
        \frac{1}{is_{n2}-s_{i1}^\prime}
        \frac{\partial^\to}{\partial s_{i1}^\prime}
        M_{mi}(s_1^\prime,s_1)\right)\cr
 &&     -\frac{1}{2}\sum_{k\neq l}
        \frac{1}{(is_{m2}-s_{k1}^\prime)
        (is_{n2}-s_{k1}^\prime)(s_{k1}^\prime-s_{l1}^\prime)^2}
        \frac{\partial^\to}{\partial s_{k1}}\cr
 &&     \left.+\frac{1}{2}\sum_{k\neq l}
        \frac{1}{(is_{m2}-s_{k1}^\prime)(is_{n2}-s_{l1}^\prime)
        (s_{k1}^\prime-s_{l1}^\prime)^2}
        \frac{\partial^\to}{\partial s_{k1}}\right]
        f(s_1^\prime)\mu_B(s^\prime,s)d[s_1^\prime] \ . \cr
 && 
\label{lem3a}
\end{eqnarray}
The proof is similar to the one of formula~(\ref{LEM1}).  We notice
that the arrow in Eq.~(\ref{H14}) is used slightly differently than
previously.  The operator $L^\to_m(s)$ acts also on a part of
$\widetilde{L}_n(s)$. This is not consistent with the definition in
Eq.~(\ref{p41b}). However, since this is obvious where it occurs, we
still use the same arrow.  We can now translate the left hand side of
Eq.~(\ref{H12}) into an expression in terms of $\Phi_4^{(1)}(-is,r)$.
After some further manipulations involving the identities in
Eqs.~(\ref{H11}), (\ref{H4a}) and (\ref{H5}) we arrive at
\begin{eqnarray}
\lefteqn{\Phi_{44}(-is,r)= \widehat{G}_{44} \exp\left(\tr r_2s_2 +
                   r_{11}\tr s_1\right)
                   \prod_{i,j}^2 R_{ji}\prod_{k=1}^4 S_{ki}}\cr
              &  & \left[\left(\frac{1}{\Delta_2^2(ir_2)\Delta_2^2(is_2)}+
                   \frac{1}{\Delta_2^3(ir_2)\Delta_2^3(is_2)}\right)\right.\cr
              &  & \left(8 R_{12}R_{22}
                     - 4 R_{22}\sum_{k=1}^4 S_{k2}^{-1}-4 L_2(s)\right)
                   \left(8 R_{11}R_{21}-4 R_{21} 
                   \sum_{k=1}^4 S_{k1}^{-1}
                   -4 L_1(s)\right) \cr
              &   &-\frac{8}{\Delta_2^3(ir_2)\Delta_2^4(is_2)}
                   \left( \trg r 
                   -\sum_{i=1}^4 S_{i1}^{-1}\right)
                   \left(L_1(s)-L_2(s)\right) \cr
              &  & \left.-\frac{16}{\Delta_2^3(ir_2)\Delta_2^3(is_2)}
                   \sum_{k=1}^4\prod_{j=1}^2
                   R_{2j}S_{kj}^{-1}\right]
                   \exp\left(-r_{11}\tr s_1^\prime\right)
                   \Phi_3^{(1)}(s_1^\prime,\widetilde{r}_1)\cr
              &   &\qquad\qquad\qquad 
                \left(ir_{12}\longleftrightarrow ir_{22}\right)\ .
\label{H12}
\end{eqnarray}
After rearranging terms this yields the result~(\ref{p413}) for
$\Phi_{44}(-is,r)$.

\end{document}